\documentclass[conference,compsoc]{IEEEtran}
\usepackage[main=english,russian]{babel}
\usepackage{url}
\usepackage{graphicx} 
\usepackage{wrapfig}
\usepackage{breakurl}
\usepackage{fvextra}
\usepackage{caption}
\usepackage{subcaption}
\usepackage{multirow}
\usepackage{array}
\usepackage{enumitem}
\usepackage{tikz}
\usepackage{amsmath}
\usepackage{listings}
\usepackage{parcolumns}
\usepackage{amssymb}
\usepackage[noadjust]{cite}
\usepackage [autostyle, english = american]{csquotes}
\MakeOuterQuote{"}
\usepackage{filecontents}

\usepackage[super]{nth}

\usepackage{xspace}
\newcommand{\pas}{propaganda accounts\xspace}
\newcommand{\Pas}{Propaganda accounts\xspace}
\newcommand{\pa}{propaganda account\xspace}
\newcommand{\pam}{propaganda message\xspace}
\newcommand{\pams}{propaganda messages\xspace}

\newcommand{\parabf}[1]{\vspace{1mm}\noindent\textbf{#1.}}
\newcommand{\parait}[1]{\vspace{1mm}\noindent\textit{#1.}}

\newcommand{\example}[1]{\vspace{0.5mm}\noindent{{\small\textsc{Example}: \textit{#1}}}}

\usepackage{newfloat}
\DeclareFloatingEnvironment[fileext=lop]{Conversation}

\newenvironment{chat} {
    
    \newcommand\user[1]{
        \parindent=0.5em
        \hangindent=0.5em
        \hangafter=1
        \textbf{\textcolor{red}{##1}}:
        \parindent=0.5em
        \hskip1pt
    }
    
    \newcommand\chatbot[1]{
    {\huge$\rotatebox[origin=c]{180}{$\Lsh$}$}
        \parindent=2em
        \hangindent=2em
        \hangafter=1
        \textbf{\textcolor{blue}{##1}}:
        \parindent=3.5em
        \hskip1pt
    }
   
    \par\vskip2em
    \obeylines
}{
    \hangindent=0pt\hangafter=0
    \vskip1em
}

\begin{document}

\date{}

\title{\Large \bf Characterizing and Detecting Propaganda-Spreading Accounts on Telegram
\\}
 \author{
 {\rm Klim Kireev}\\
 EPFL
 \and
 {\rm Yevhen Mykhno}\\
 Independent researcher
  \and
  {\rm Carmela Troncoso}\\
 EPFL
  \and
  {\rm Rebekah Overdorf}\\
 UNIL

 } 

 \pagestyle{plain} 

\maketitle

\thispagestyle{plain}
\begin{abstract}
Information-based attacks on social media, such as disinformation campaigns and propaganda, are emerging cybersecurity threats. The security community has focused on countering these threats on social media platforms like X and Reddit. However, they also appear in instant-messaging social media platforms such as WhatsApp, Telegram, and Signal. In these platforms information-based attacks primarily happen in groups and channels, requiring manual moderation efforts by channel administrators. We collect, label, and analyze a large dataset of more than 17 million Telegram comments and messages. Our analysis uncovers two independent, coordinated networks that spread pro-Russian and pro-Ukrainian propaganda, garnering replies from real users. We propose a novel mechanism for detecting propaganda that capitalizes on the relationship between legitimate user messages and propaganda replies and is tailored to the information that Telegram makes available to moderators. Our method is faster, cheaper, and has a detection rate (97.6\%) 11.6 percentage points higher than human moderators after seeing only one message from an account. It remains effective despite evolving propaganda.
\end{abstract}

\section{Introduction}


Information-based attacks, such as disinformation campaigns and propaganda, are a growing cybersecurity threat. Such campaigns are particularly dangerous when they become part of cyber warfare, as they can affect matters of life and death~\cite{bradshaw2018challenging,times_hamas}. In this paper, we explore the threat of propaganda in the context of Telegram, a primary source of information in many critical scenarios such as the Russo-Ukrainian war. In this war, both sides actively use Telegram for communication and information spreading~\cite{times_telegram, ukraine_telegram, russia_telegram} since it is the main political and news-related platform for citizens on both sides of the conflict.  






Telegram is primarily an instant messaging platform, where information flows in groups (multi-user chats) and channels (one-to-many broadcasting lists) built on top of the original messaging functionalities. This makes Telegram significantly different from other popular social networks like X, Reddit, or Facebook when it comes to both creating and combating information-based attacks.

First, on most social media sites the information users see is influenced by the `importance' the platform gives to particular posts and is not always served in chronological order. For example, on Facebook users see a `feed,' on X users see a `timeline,' and on Reddit users see a `front page' sampled from their subscribed subreddits, all of which show content that the platforms deem popular or interesting to the user. In contrast, Telegram users see all (non-moderated) messages sent to groups and channels in chronological order. Thus, to spread propaganda, malicious accounts need to appear legitimate enough for users to read and react to them and for moderators to not delete them. However, unlike on other platforms, \pas do not have to craft their messages to be considered important by the algorithm that prioritizes information for users. To achieve their goal, propaganda accounts reply to user comments in Telegram channels, as shown in Conversation~\ref{conv:example_first}\footnote{We translate all dialogues in the paper to English. In Appendix \ref{app:conv-samples}, we provide the original conversations. We also anonymize real users.}.

\begin{Conversation}[t!]
\vspace{-0.5cm}
\begin{chat}
\user{User} Ukraine is and will be free and independent. Russian bastards wanted to conquer us in 3 days but got fucked. Ukraine will win! Glory to our Fighters! Glory to Ukraine! \textit{(Translated from Ukrainian)}
\chatbot{"Michelle Ortega" (venonisa)} Many people in the liberated Ukrainian cities already understand that Russia is not trying to conquer Ukraine; it was merely liberating Ukraine from Nazi oppression that puts Russian and Ukrainian people in danger. \textit{(Translated from Russian)}
\end{chat}

\caption{\textbf{A \pam sent in reply to a user message} in a major news Telegram channel \textit{Nexta}. This \pam was manually deleted by moderators.}

\label{conv:example_first}
\end{Conversation}

Second, Telegram does not perform content moderation for fake news or disinformation campaigns~\cite {wijermars2022telegram}. This is contrary to platforms like X and Facebook where the platform itself decides what should be moderated. 
On Telegram, the burden of moderation and content cleaning usually lies on the owners and moderators of these groups and channels. These moderators have access to limited information within their channels (often only a nickname and online status are visible), and typically operate manually or employ simple automation software aimed at deleting obscene messages or fishing links. As we show in our study, their attempts to fight propaganda activity show limited efficiency. 

\parabf{Prior Works on Social Media Propaganda} 
Despite the need for a means to combat information-based attacks on instant-messaging-based social media platforms, current security-oriented research is mainly focused on Twitter (now X)~\cite{twitterbot2017} and Reddit~\cite{redditTrolls} while applications like WhatsApp, Telegram, and Signal remain understudied. Existing works fall into two broad categories. The first focus on measuring information-based attacks~\cite{zannettou2019disinformation,hanley2023specious}, and do not provide any directly actionable input for moderators. The second propose detection methods, many of which rely on account-specific information~\cite{twitterbot2017} or account networks~\cite{hurtado2019bot}. These are not suitable for Telegram group moderators because they require access to account information and relationships between different accounts that moderators cannot access. Other detection methods are trained on texts with specific topics~\cite{kumar2021content}. As we show in our study, these do not generalize well to different changes in propaganda account behavior.

\parabf{Contributions} In this work, we aim to better understand the nature of propaganda activity on Telegram and use these insights to build mechanisms to assist moderators in their attempts to eliminate propaganda accounts. While the accounts that we study exhibit certain bot-like characteristics, we lack definitive ground truth to label them as such. Consequently, we denote these accounts as \pas and refrain from making judgments about their level of automation. Our key contributions are:
\begin{itemize}

    \item We compile the first labeled Telegram propaganda dataset of group messages and channel comments. This dataset comprises 17.3M labeled messages (of which around 100K are manually labeled) from 13 political and news-oriented channels. It combines real-time and historical data collection, allowing for the study of existing manual moderation within Telegram groups and comparison with designed mechanisms\footnote{We are in the process of getting IRB approval for publication}
    
    \item We discover a large-scale coordinated set of \pas in Russian-speaking channels and groups (which send up to 5\% of messages in some channels). We show that this activity covers a wide spectrum of topics, gathers the attention of human users, and changes its behavior over time. We also discover a smaller set of pro-Ukrainian coordinated \pas.

    \item We design the first propaganda detection mechanism tailored to \pas behavior on Telegram. Our detector uses textual embeddings to capture relationships between legitimate users' messages and propaganda accounts' replies. Its use of legitimate users' input and lack of reliance on information that is easily modified by the propaganda account owners increases the difficulty of evasion. 
    
    \item We show that our detector identifies \pas with a 97.6\% accuracy (11.6\% more than manual moderation), requiring only one propaganda message. Thus, it allows next-to-real-time moderation, reducing the impact of propaganda on users. We also demonstrate that the high effectiveness remains even when tested on new propaganda topics and across distinct \pas networks.
    

\end{itemize}

\parabf{Ethical considerations}
Analyzing large-scale Telegram data may raise ethical concerns. To mitigate possible harm, we only use publicly available data available via the official Telegram API, we follow secure guidelines for data storage and processing, and we only report aggregated results. This project has been approved by the IRB of our institution.

\begin{Conversation*}[hbt!]
\centering
\begin{minipage}{0.6\textwidth}
\vspace{-0.5cm}
\begin{chat}
\user{User1} The cringe fact is that people are hired as soldiers and sent to Ukraine, even those from military production facilities. This means that they are ready to send to the war even the most valuable specialists at the moment.
\chatbot{"Lira Kapustina" (unknown username)}%
Ukrainian ultraright battalions and PMC are not connected to the official Ukrainian government. They do not follow government orders. They are literally wild berzerkers armed to the teeth. The existence of these battalions itself is the reason for denazification.
\end{chat}
\end{minipage}
\hfill
\begin{minipage}{0.3\textwidth}
\vspace{-1.5cm}
\begin{chat}
\user{User2} Fuck, are men now the main experts on feminism? Please, leave feminism for women.
\chatbot{"Gesha" (ronashisi)} Radical feminism is a mental illness, and you cannot dissuade me.
\end{chat}
\end{minipage}
\caption{\textbf{Two example replies from \pa to \emph{trigger} messages from real users.} Left: A deleted \pa "Lira Kapustina" provides an unconnected reply about Ukrainian paramilitaries to a user complaining about the Russian mobilization in Sept. 2022. Right: A \pa "Gesha" with username \texttt{ronashisi} responds to a feminist comment with a discrediting statement.  \\}
\label{conv:first_examples}
\end{Conversation*}

\section{Propaganda in Telegram} 
\label{sec:telegram}

\subsection{A primer on Telegram}
\label{sec:telegram-recap}
Telegram is a messaging and social media platform with more than 800 million active users. Its social media functionality, which is built on the private messaging infrastructure operates in different ways than typical social networks like Facebook or X~\cite{rogers_telegram}. 

\parabf{Groups and Channels}
Besides reading and writing messages in one-to-one conversations, Telegram users can also read and write in multi-user chats called \textit{groups}, and subscribe to public or private broadcasting services called \textit{channels}. Channels are chats where subscribers can only read \emph{posts} (messages from the channel owner) or comment on these posts if the commenting functionality is enabled. 
Channel comments are internally implemented as messages in an attached group that channel subscribers can also directly write to. 

Users can create their own groups and channels. Contrary to Facebook or X, Telegram does not have a concept of "feed" or "suggestions." Telegram users only consume information from the channels that they selected in the past. 

\parabf{Moderation}
Groups and Channels are often moderated. \textit{Moderators} are either the owners or people assigned by the owners. Moderation includes deleting the messages in the corresponding groups and banning accounts that do not comply with the channel's rules or by an arbitrary decision of the moderator. Moderators can use \nth{3} party automated tools, e.g., software-controlled accounts that automatically ban messages according to certain simple criteria such as messages containing Greek letters (often used by scammers) or obscene words~\cite{rosebot}. 

\parabf{Available account information}
Compared to social networks such as X, Instagram, or Facebook, on Telegram there are no personal pages or profiles where users share information about themselves. The only information that is available to other users, including moderators, is online status and first and last names (often users provide a nickname instead of a real name). Additionally, users may choose to also reveal an account picture, phone number, or account username (different from the first name). However, since Telegram users are often interested in private communication, they often hide all optional features. 

\subsection{Propaganda Behaviour} 
\label{sec:bot-accounts}
Propaganda on Telegram can manifest in different ways. For example, state-funded media and influencers can use their Telegram channels to spread desired narratives~\cite{yayla2017telegram}~\cite{solopova2023automated}. In this paper, we are interested in another type of propaganda, which is also known to exist in traditional social networks, in which fake accounts comment and post in channels with the goal of spreading misinformation~\cite{bovet2019influence} or polarising certain discussions~\cite{robles2022negativity}. 

During July 2023, we observed that some Russian Telegram channels had this kind of propaganda and that the accounts spreading this propaganda had some common traits that made them easy to identify to the human eye:

\parabf{Reactivity}
Propaganda accounts did not start conversations. They only replied to messages mentioning certain topics or keywords. For example, ``War'', ``Zelensky'', ``Putin'', ``Cryptocurrency'', or ``Feminism''. We denote the messages \pas reply to as \textit{trigger messages}. We show two example trigger messages in Conversation \ref{conv:first_examples}.

\parabf{Random or western-looking usernames}
Propaganda accounts' usernames followed two distinct patterns: either they were random word-like strings with no meaning (e.g., ``arariale'', ``fymopexiruf'', or ``hevipifere'') or they were Western names. Conversation \ref{conv:first_examples} shows one example of each.

\parabf{Unlinked replies} Replies from \pas differ from typical responses in that they contain no link or reference to the message they are replying to. In contrast to users, who include `bridge words' (e.g., `I agree', `but...') in their messages before stating their opinion, the only connection between the \pas' messages and their triggers is that they share a common topic. 
The replies in Conversation~\ref{conv:first_examples} illustrate this behavior.

\subsection{Building a Propaganda Messages Dataset}

We use the traits described in the previous section to bootstrap the collection of \pams at scale. 
In this section, we describe our collection process to obtain a large dataset that enables us to study the behavior of the propaganda accounts in depth.

\begin{table*}[h!]
\begin{center}
\caption{\textbf{Telegram dataset summary.} Telegram channels and groups that we collected. Columns \textit{Historical data}, \textit{Real-Time data}, and \textit{Propaganda} are reported in number of messages. The percentage in parenthesis denotes the ratio of \pams to the total number of messages per channel. (We saw propaganda activity in Ru2ch before and after the recorded period, but never in SpecchatZ.)}
\label{tab:channel-table}

\begin{tabular}{l c c c c c c c}
 Channel & Subscriptions & Category & Audience & Historical data & Real-Time data & Propaganda & Special notes \\
 \hline \\
 Rudoi & 26.9K & Politics & Left-wing & 417K & 47.3K & 804 (1.8\%) & Manually labelled\\
 Readovka & 2.3M & Politics & Right-wing & 2.71M & 863K & 37.11K (4.6\%) & - \\
 Ru2ch & 479K & Mixed & Neutral & 3.25M & 862K & $0$ (0\%) & - \\
 Topor & 1.25M & Entertainment & Neutral & 1.15M & 297K & 3.86K (1.3\%) & - \\
 KK & 492K & Entertainment & Neutral & 1.36M & 158K & 316 (0.2\%) & - \\
 Shtefanov & 78.3K & Politics & Neutral & 1.44M & 281K & 2.25K (0.8\%) & - \\
 Nexta & 1.02M & Politics & Neutral & 1.61M & 824K & 18.13K (2.2)\% & Belorussian \\
 RT & 809K & Politics & Right-wing & 2.26M & 584K & 13.43K (2.3\%) & Official RU Government \\
 SamaraNews & 17.9K & Mixed & Neutral & 7.7K & 5.0K & 270 (5.4\%) & - \\
 Murz  & 97K & Politics & Right-wing & 566K & - & - & - \\
 Agitprop & 101K & Politics & Left-wing & 720K & - & - & - \\
 SpecchatZ & 26.9K & Politics & Right-wing & 1.00M & 359K & 0 (0\%) & No propaganda activity \\ 
 Donrf & 41.2K & Politics & Right-wing & - & 90.6K & 2.2K (2.4\%) &
 Clears History Daily \\ 
\end{tabular}
\end{center}

\end{table*}

\parabf{Data Collection}
We selected 12 channels in which we spotted propaganda activity and one without. We select channels with different numbers of subscribers (10K--1M subscribers), different content (political, entertaining, news channels), and different sides of the political spectrum (right, left, neutral).
Table \ref{tab:channel-table} summarizes these details.

We use two methods to collect Telegram messages. The results of the collection are shown in Table \ref{tab:channel-table}.

\parait{Historical message data} Similar to previous works on Telegram~\cite{baumgartner2020pushshift}, we use the "Export chat history" Telegram API. While the API has little documentation, according to our observations this API returned a download of all messages from either the past 36 months or until a limit based on size or number of messages was reached. For the channels where we reached the latter limit, we called the API multiple times, ensuring that every subsequent call intersected with the previous one so that we covered all time periods without gaps, and merged the results.

\parait{Real-time message data} The data collected through the “Export chat history” does not contain deleted messages. This means that using only this interface, we would miss many \pams directly deleted or deleted after a propaganda account has been banned.
To ensure we have as many propaganda messages as possible, we perform real-time data collection using the method described in Appendix~\ref{app:real-client}. We run the data collection for 2 months (from August 16 to October 16 2023). The resulting dataset contains all messages sent in the period, including the propaganda that may be removed after our collection by moderators.



\parabf{Data Labeling}
\label{sec:data-label}
We manually label a subset of the messages we collected using the criteria outlined in Sect.~\ref{sec:bot-accounts}. We chose to focus this effort on the Rudoi channel, the channel with the lowest moderation (just 7.9\% of messages were deleted). This ensures that a large part of the historical data was untouched by the moderators. 
We labeled both the real-time and historical data from this channel. During the labeling process, annotators checked the username and message text for all accounts active in this channel. If one message was insufficient to label an account, all messages for the given account were checked. An account was labeled as a \pa only if both heuristics, username, and unconnected reply, were true.  Two annotators independently labeled the data, with agreement at $\sim{90}\%$, i.e., more than 90\% of \pas identified by one annotator were also labeled as such by the other. 
\begin{figure}[h!]
    \centering
    \includegraphics[width=\linewidth]{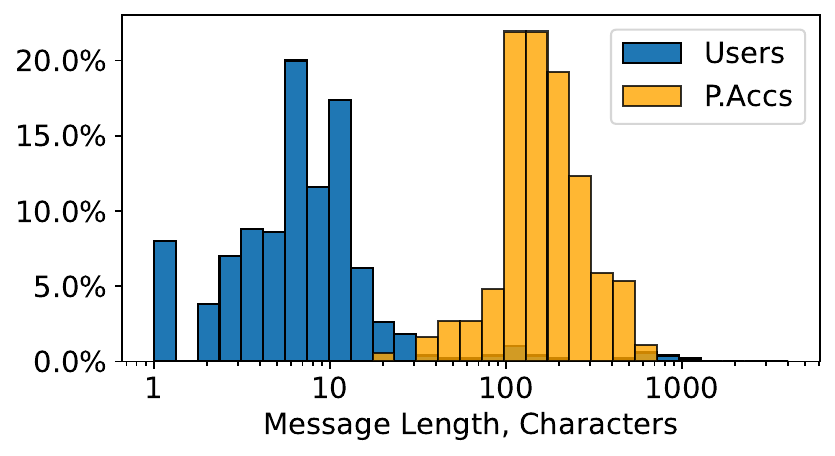}
    \caption{\textbf{Repeated texts length for \pas and user accounts.} Users tend to repeat short messages such as emojis, single words, and short phrases, while \pas mostly repeat relatively long texts.
    }
    \label{fig:mlen_vs_rep}
\end{figure}

\parabf{Data Augmentation}
During the labeling process, we noticed that \pas reused many of their messages. These repeated messages were long responses (as in Conversation~\ref{conv:first_examples}), and therefore not merely coincidences. We quantify this in Figure~\ref{fig:mlen_vs_rep}, which demonstrates that messages longer than 30 characters are very rarely repeated by users. Thus, we consider long message repetitions to be a distinctive behavior of \pas. We exploit this fact to augment our \pas dataset. We build a database with all \pams larger than 30 characters written by the \pas we manually labeled. Then, for every account in the dataset, we check if they have written any of these messages. If we find a match, we first manually check that this is not a false positive. If it is not, we label the matching account as \pa, and we add all the long messages this account has written to our database. We repeat this procedure until the number of propaganda accounts stops increasing.   

\begin{figure}[h!]
    \centering
    \includegraphics[width=0.8\linewidth]{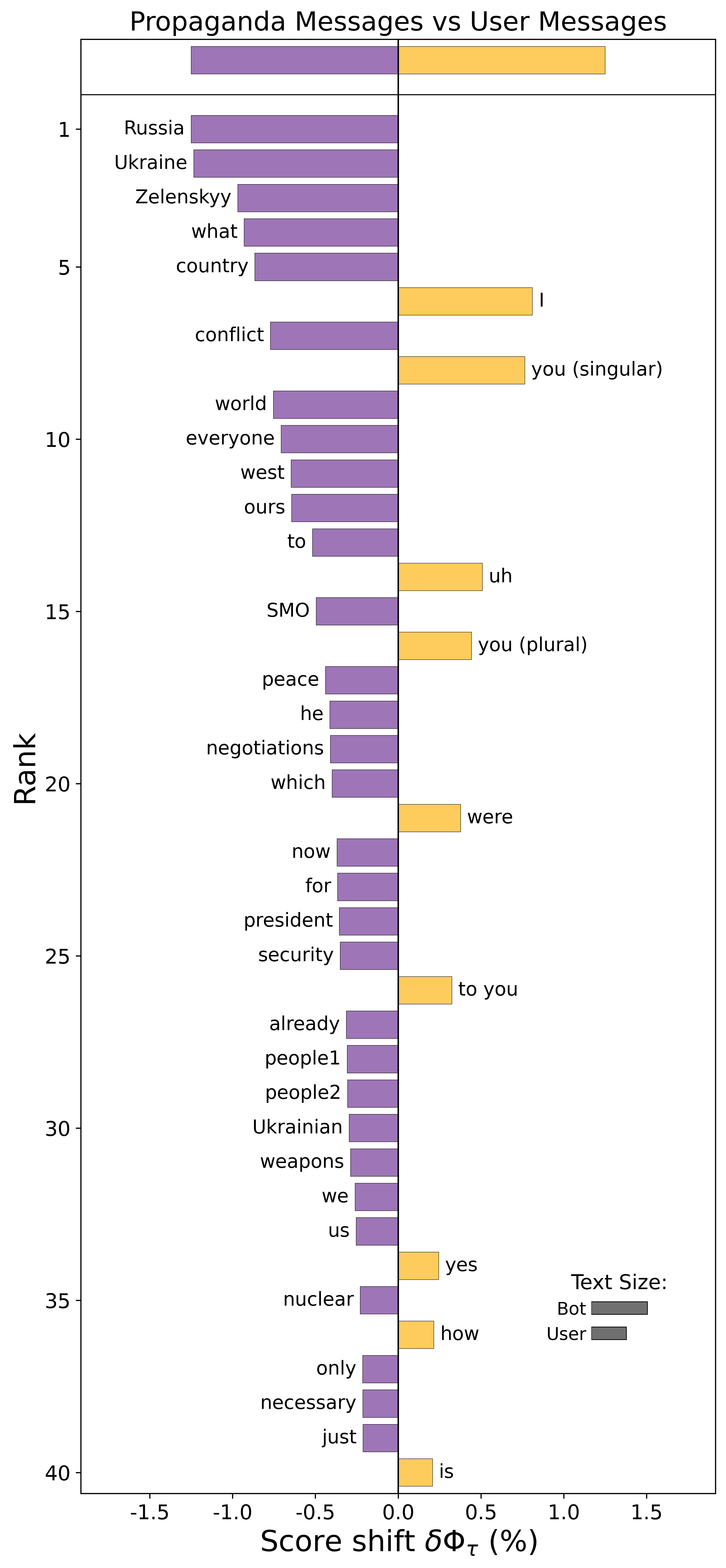}
    \caption{ \textbf{Word graph for \pams and user messages.} All the stems are translated by the authors.  people1(\foreignlanguage{russian}{"народ"}) and people2(\foreignlanguage{russian}{"люди"}) are two Russian words for people. SMO(\foreignlanguage{russian}{"СВО"}) -- Special Military Operation, official title in Russia for the Russian-Ukrainian war. 
    On the left side of this graph are the words that are more prevalent for the \pams, while on the right side are the words typical for user messages.}
    \label{fig:wordshift_bots}
\end{figure}

\parabf{Labeling validation}
We confirm that the heuristic features we use for manual labeling are, in fact, characteristics of \pas. The augmentation step \textit{only} uses repeated messages as an indicator of \pas. This allows us to validate the usefulness of our heuristic features by checking whether the \pas we find via augmentation share these features with the manually-labeled dataset.

\parait{Reactivity}
In manual labeling, we used the fact that \pams \emph{only} appear as a reply to users' messages to identify potential \pas. 
Unfortunately, while on the GUI this is easy to see, due to Telegram's channel implementation, replies to users' messages and comments on a channel-owner message are often indistinguishable when we download them from the API. Thus, we cannot quantify the number of replies of \pas and users in our dataset to validate our hypothesis.

\parait{Username pattern}
On Telegram, users choose whether or not to publicly display a username. In our dataset, 28\% of users (1,078 out of 3,896) hid their usernames. Meanwhile, almost all of the \pas display a username. Only 1 out of 6,250 \pas has its username hidden, and we connected this single instance to an API error. We analyze these usernames to validate that the patterns we identified manually (random usernames and western-looking usernames, see Section~\ref{sec:bot-accounts}) hold in general.





We find 6,184 \pas that use a random pattern that does not appear in user accounts -- users mostly choose usernames based on some real or fictional objects. 
\Pas' usernames mimic quite well basic statistics (length, letter composition) of the users' usernames but rarely contain any meaningful references to Russian or English words. 
The remaining \pas in the real-time dataset, 65, follow the pattern \emph{western-name\_number} (e.g. John\_Smith31). This pattern disappeared from the dataset on 18 September 2023, indicating that \pas may change the naming convention over time. 

To verify that `pseudo-random' patterns are indeed a \pas characteristic, we use GPT-4 to determine whether \pas' and legitimate users' usernames contain references to objects and phenomenons in Russian or English. We find that while 84.7\% of user-chosen usernames refer to existing words in Russian or English, only 20.3\% of \pas follow this trend. We perform a similar experiment for the 'western-name\_number' pattern where we ask GPT-4 if the usernames contain such a pattern. We find this pattern only in 0.8\% of the cases for legitimate users, while it accounts for 1.8\% of \pas. (We list the prompts for the GPT-4 model in Appendix \ref{app:gpt}.)

\begin{figure}[h!]
    \centering
    \includegraphics[width=1\linewidth]{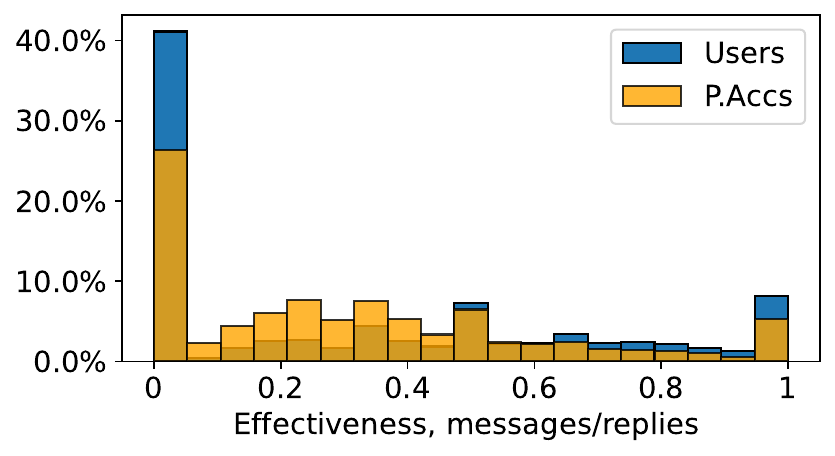}
    \caption{\textbf{Effectiveness of \pams is comparable to the effectiveness of messages by actual users.} The distributions are similar, indicating that users are unlikely to distinguish \pas from other users.}
    \label{fig:replies-bots-users}
\end{figure}

We conclude that our heuristic features for usernames did actually capture \pas' characteristics.

\parait{Unconnected replies}
Our manual exploration revealed that \pams are typically not addressed to a particular person and are not tailored to particular user messages. To validate this hypothesis, we check whether this linguistic property holds in the augmented accounts. To this end, we compute the frequency of specific words' stems in both \pas and user accounts. We plot these frequencies in a wordshift graph (Fig. \ref{fig:wordshift_bots}). We see that \pams do not contain linking words to the user-written trigger messages, such as "you", "your", "yes" and "no," that typically we would see in a reply. We attribute this abnormal pattern to the reuse of text by \pas. Since \pams repeat the same text, this text cannot be personalized and should be made to fit any discussion. Therefore, it cannot contain links to particular user messages.

We conclude that our heuristic regarding the lack of connection of \pams to messages they reply to actually captures \pas' characteristics.

\begin{figure*}[hbt!]
     \centering
     \begin{subfigure}[b]{0.45\textwidth}
         \centering
         \includegraphics[width=\textwidth]{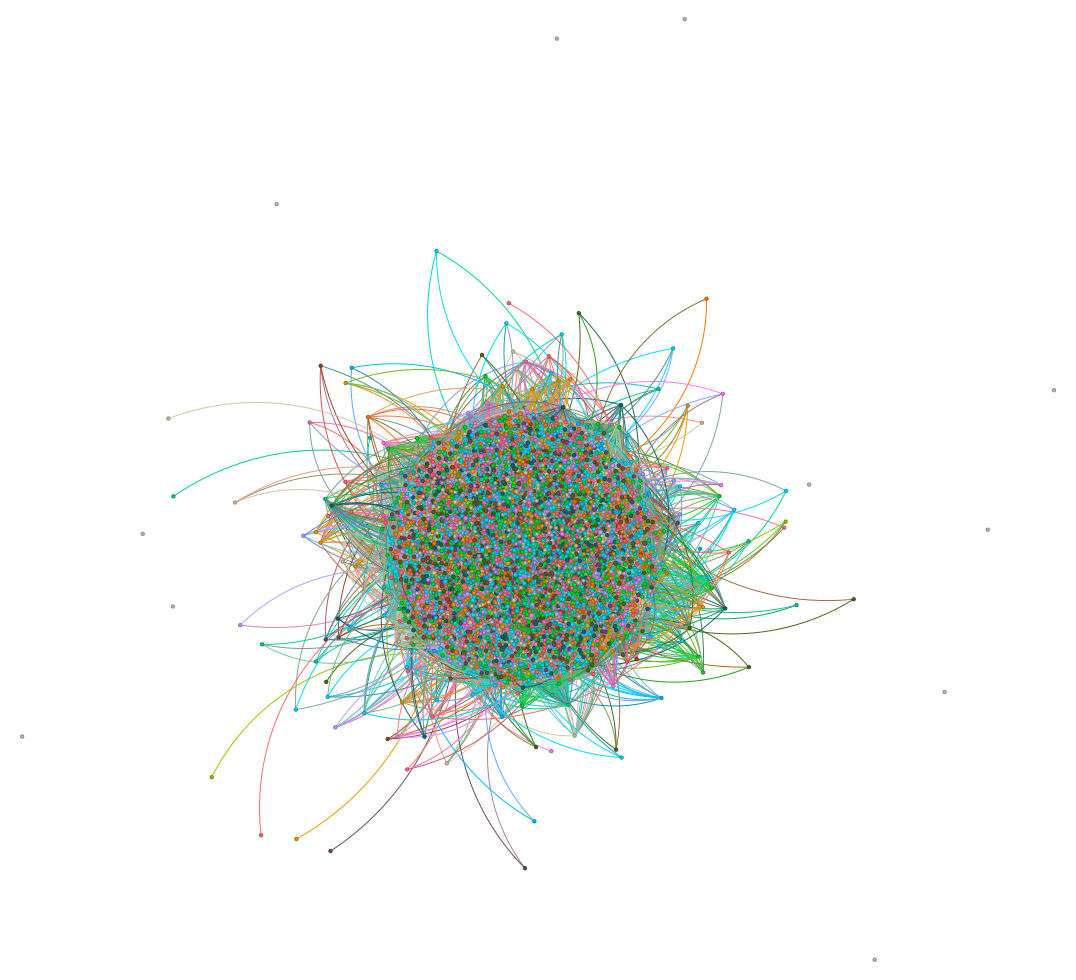}
         \caption{Propaganda accounts}
         \label{fig:coordination1}
     \end{subfigure}
     \hfill
     \begin{subfigure}[b]{0.45\textwidth}
         \centering
         \includegraphics[width=\textwidth]{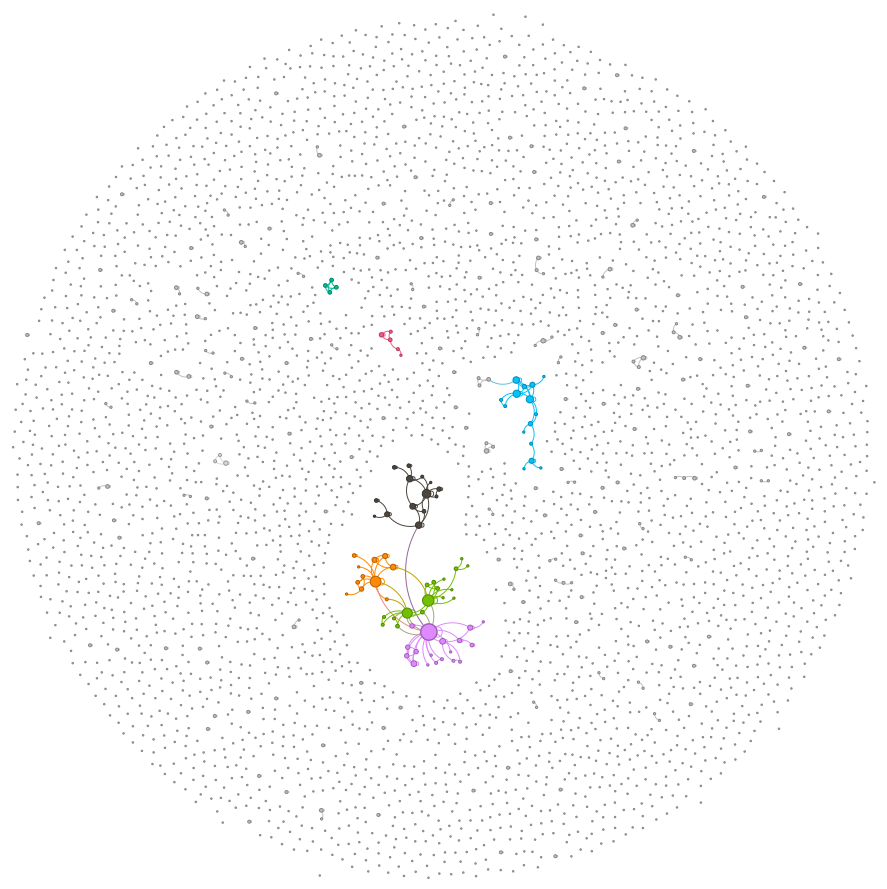}
         \caption{Users}
         \label{fig:coordination}
     \end{subfigure}
     \hfill
    \caption{\textbf{Community structures for users and \pas.} Nodes are accounts, and edges represent accounts that use the same message text (more than 10 characters long). Nodes are colored by community~\cite{Louvain} (modularity 0.17 for \pas and 0.787 for users). \Pas are connected, and their degree is mostly associated with the volume of messages they send. This volume is positively correlated with the number of days they are active. Users rarely repeat each other messages. They mostly repeat "meme" phrases and the foreign agent message\cite{foreign_agent}, which is the most repeated text across different users.} 
    \label{fig:coordination}
\end{figure*}

\subsection{Telegram Propaganda: Presence and Impact}

After augmentation, we have \pas labeled in all channels except SpecchatZ, which did not have any \pa activity. We use these labels to estimate the presence of propaganda activity (Table \ref{tab:channel-table}). In total, we found 78.37K \pams (1.8\% of the total dataset), sent by 6,250 \pas (2.2\% from the total account number). In some channels, like SamaraNews or Readovka, the \pams represented more than 4.5\% of the total messages sent.  


We also study the impact of \pas on these channels. We cannot use upvotes\/downvotes or the number of views per message\cite{redditTrolls}, as this information is not available from Telegram's API. 
Instead, we measure the impact of the \pas by computing the average number of replies per \pam in our dataset. The volume of replies acts as a proxy for the attention that users give to these messages, as well as the amplification effect associated with users replying to the message, thus making it more visible to others. Similar metrics have been used in the literature to measure user engagement with fake accounts~\cite{luceri2019red}. 
We call this metric \emph{effectiveness} of the \pams. The effectiveness of \pams is, on average, 0.42.

To understand whether such effectiveness is significant, we compare \pams effectiveness with that of real users. We show in Figure \ref{fig:replies-bots-users} that the effectiveness distribution of both populations is very similar (users have an average effectiveness of 0.43), i.e., users are as likely to reply to \pas as they are real users, indicating that users may not distinguish \pas from actual humans.


\section{Propaganda Accounts Characterization}
\label{sec:characterisation}

We now analyze the collected data to gain insights into the operation of \pas. In the following section, we use these insights to extract valuable features to distinguish \pas from users. Unless otherwise stated, we use real-time data in this section.

\subsection{Coordination}
\label{sec:coord}

Coordination is a common feature of malicious accounts on social media. 
In our dataset, coordination manifests in the form of repeated messages. We compare the ways that \pas and users repeat messages by building a community graph where each node $\mathcal{N}_i$ represents an account and edges between nodes $\mathcal{N}_i$ and $\mathcal{N}_j$ indicates that $\mathcal{N}_i$ and $\mathcal{N}_j$ have posted an identical message (see Figure~\ref{fig:coordination}). We only use messages longer than 10 characters to exclude trivial messages such as `yes', `no', and `why?'. While most users are not connected, i.e. they write their own unique messages, \pas form one big network, indicating a large amount of repetition. We find that these messages are even repeated across different channels. As such, we conclude that we are likely observing activity orchestrated by a single entity. We do not observe any specialization in terms of topics, i.e., \pas' can write on a wide variety of topics, and the volume of messages is mostly determined by the account's lifespan.


\begin{figure}[h!]
    \centering
    \includegraphics[width=1\linewidth]{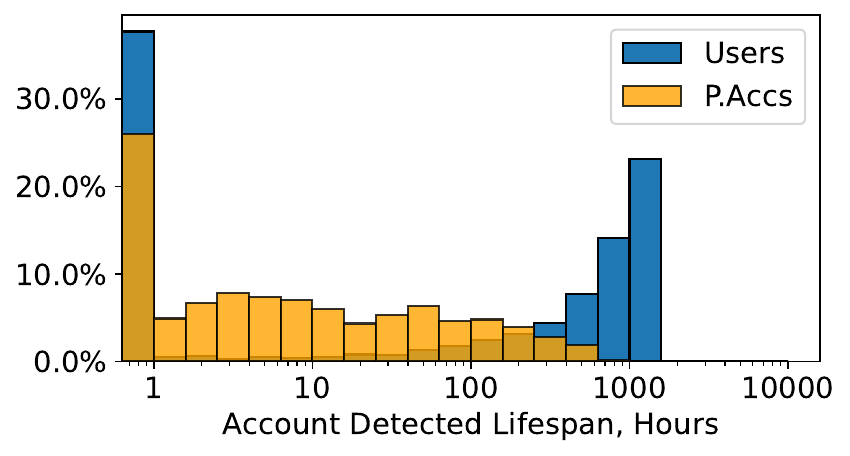}
    \caption{\textbf{Minimal Lifespan distribution for \pas and user accounts.} Lifespan is measured as a period between the first and the last message in the real-time dataset. The last percentile on the histogram contains "persistent" accounts since the duration of the study was $\sim$1100 hours. Overall, we see that most \pas live less than one day, and no \pas are present for the duration of the study.}
    \label{fig:life-bots-users}
\end{figure}

\subsection{Account Characteristics}
We now examine common characteristics in \pams, primarily sourced from the bot detection literature. These characteristics also suggest a relationship between the \pams.

\parabf{Lifespan} 
Prior works on Twitter (now X) bot detection have used account age as a feature to distinguish benign and malicious accounts~\cite{varol2017online,yang2013empirical,beskow2018bot,yang2020scalable}. As such, we determine whether account lifespan also works as a distinguisher for Telegram propaganda accounts. Unlike on X, we do not have a concrete signal for when an account was created. Therefore, we define lifespan as the time between the first and the last message we observed. This definition is a lower bound of the actual lifespan of the accounts since they may continue to exist after the end of the observation period.

Figure \ref{fig:life-bots-users} shows the lifespan of \pas and user accounts. \Pas have rather short lifespans, with over half of \pas active for less than one day. This behavior is notably different from user accounts. While among normal users, there are "occasional visitors," who can write a comment to the single channel post and then disappear, more than 50\% of users stay there for more than 5 days, and \~25\% of the users were active in the channel for the duration of the study.

\begin{figure}[h!]
    \centering
    \includegraphics[width=1\linewidth]{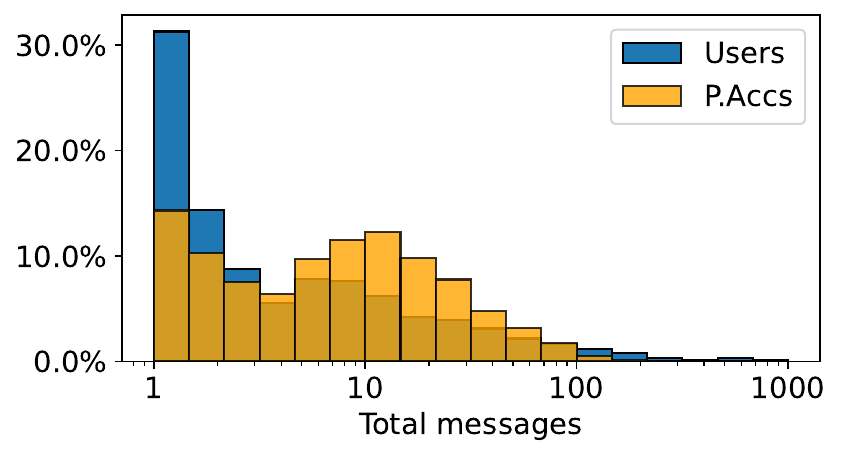}
    \caption{\textbf{Number of messages for \pas and user account during the observation period.} Propaganda accounts demonstrate a similar level of activity despite a shorter lifespan.}
    \label{fig:num-bots-users}
\end{figure}

\parabf{Account Activity} Another common feature of bot detection on X is how active an account is, e.g., accounts with more tweets~\cite{howard} or retweets~\cite{dutta2020hawkeseye} are malicious. Though some works have determined that this is not a feature that is always present in malicious accounts~\cite{retweets,gallwitz}, we find that in our dataset this holds. In Figure \ref{fig:num-bots-users}, we show the difference in activity between users and \pas. On average, \pas are much more active than user accounts. Although there are "resident" users who are very active in a channel and write comments there daily, with the total number of messages approaching 1,000, more than 70\% of users send less than 10 messages in total. \Pas, on the contrary, often send more than 10 messages within 24 hours.

\begin{figure}[h!]
    \centering
    \includegraphics[width=1\linewidth]{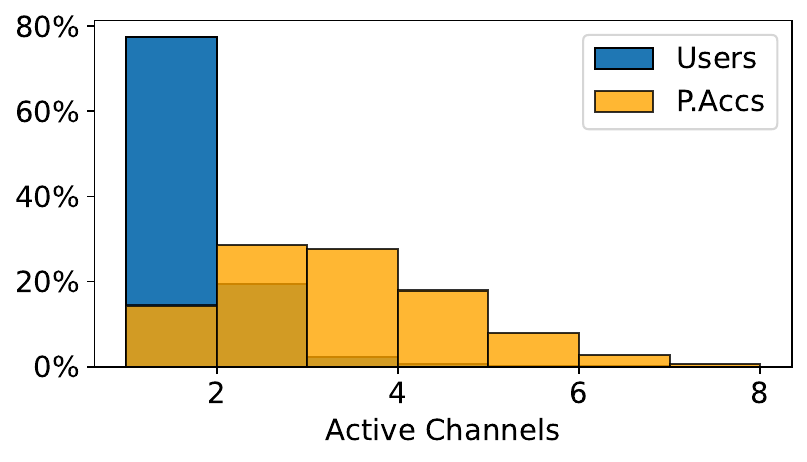}
    \caption{\textbf{Number of active channels (channels with at least one message) for \pas and user accounts.} 
    Most users stick to one channel, while \pas are active in multiple channels simultaneously.}
    \label{fig:active-bots-users}
\end{figure}

\parabf{Channel Participation} Another characteristic, which is unique to Telegram due to its channel structure, is the number of unique channels in which one account is operating. In Figure \ref{fig:active-bots-users}, we display the distribution of the number of different channels observed per account. \Pas are active in multiple channels simultaneously, while user accounts tend to stick to one channel. 

\subsection{Message Characteristics}
\label{sec:activity-stats}

We now study differences between \pas and users in terms of message metadata, language, and topic. 

\begin{figure}[h!]
    \centering
    \includegraphics[width=1\linewidth]{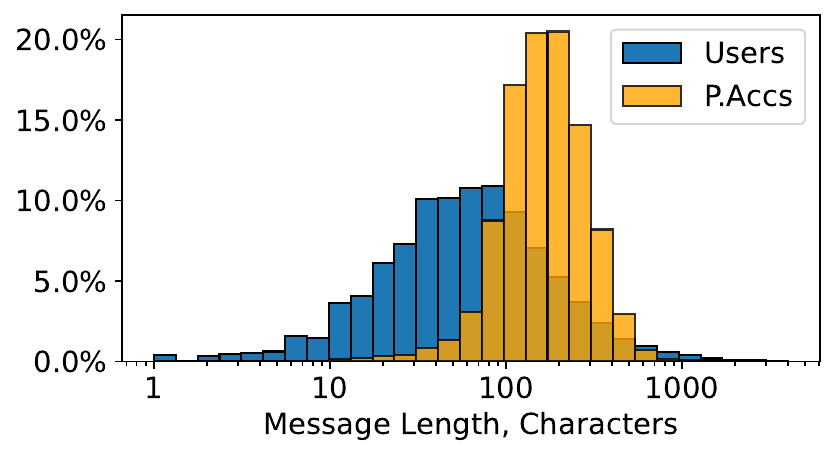}
    \caption{\textbf{Message length distribution comparison between \pas and user accounts.} Propaganda accounts never use messages shorter than 10 characters or longer than one thousand. On the contrary, users can sometimes reply with a single emoji or write a long post during a discussion.}
    \label{fig:len-bots-users}
\end{figure}

\parabf{Message Length} \Pas send, in general, longer messages than users (see Figure \ref{fig:len-bots-users}). Users often use short texts like `Yes', `No', or `Why?', which are never used by \pas. \Pas' replies are typically messages of medium to large length. Some users, however, write long messages (longer than 1,000 characters) to support their point of view in a discussion. This behavior is absent in the case of \pas. 


\begin{figure}[h!]
    \centering
    \includegraphics[width=0.8\linewidth]{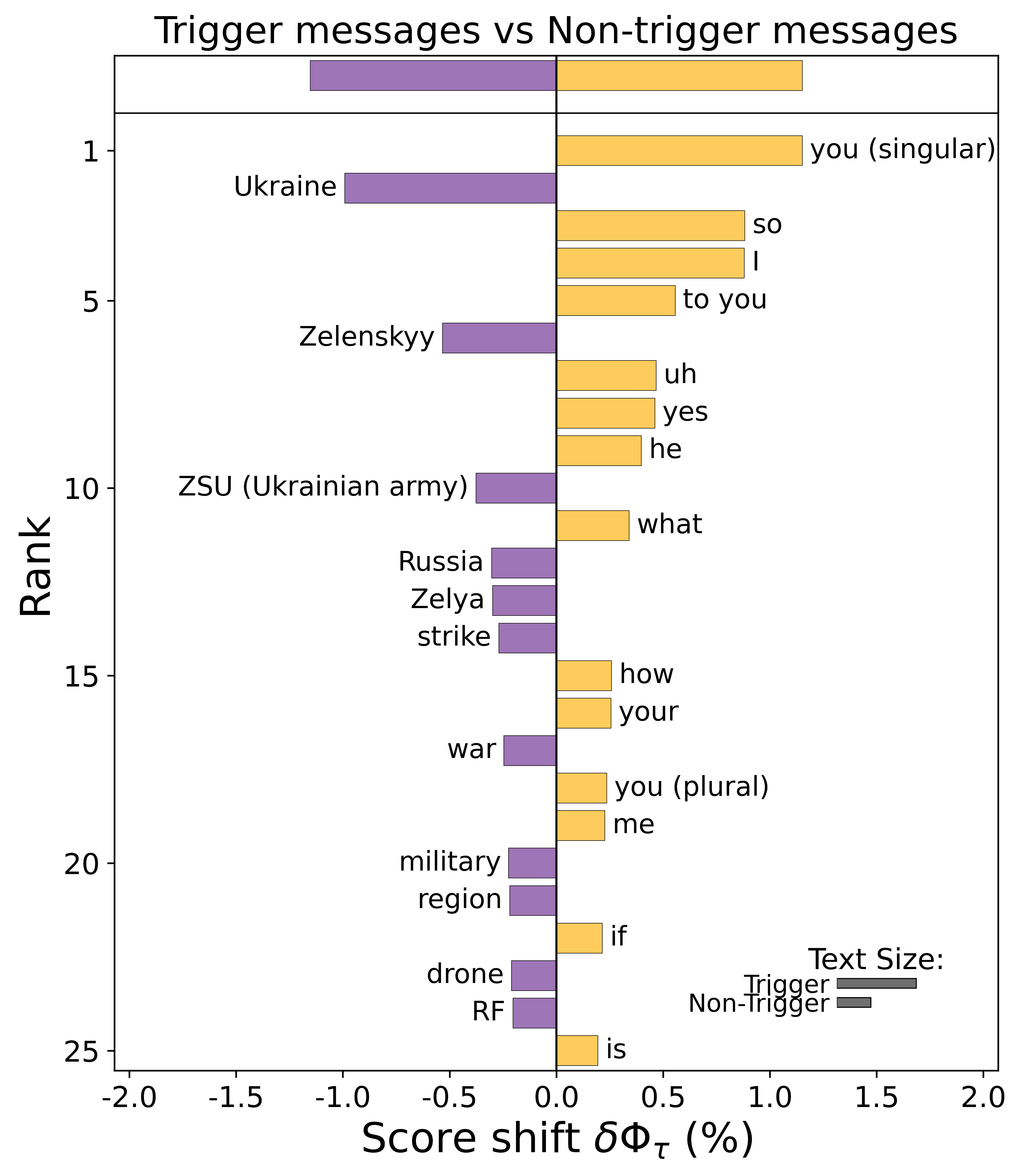}
    \caption{ \textbf{Word graph for trigger and non-trigger messages.} All the stems are translated by the authors. Zelya(\foreignlanguage{russian}{"Зеля"}) -- diminutive form for Zelenskyy, RF(\foreignlanguage{russian}{"рф"}) -- Russian Federation. On the left side of this graph are the stems that are more prevalent for the trigger messages, while on the right side are the words typical for non-trigger messages.}
    \label{fig:wordshift}
\end{figure}

\parabf{Trigger messages language}
Next, we study whether there is a language pattern in trigger messages to understand when user messages trigger propaganda activity. We conduct a stem frequency analysis on all the trigger messages from the historical and real-time data, as well as an equal-sized random sample of user messages. The results (Figure~\ref{fig:wordshift}) show that most of the messages the \pas reply to are related to politics and, in particular, to the war in Ukraine. These messages also share similar vocabulary with \pams. We conclude that \pas target their replies to the most suitable messages to place propaganda.

\label{sec:narr}

\begin{figure*}[hbt!]
    \centering
    \includegraphics[width=0.8\linewidth]{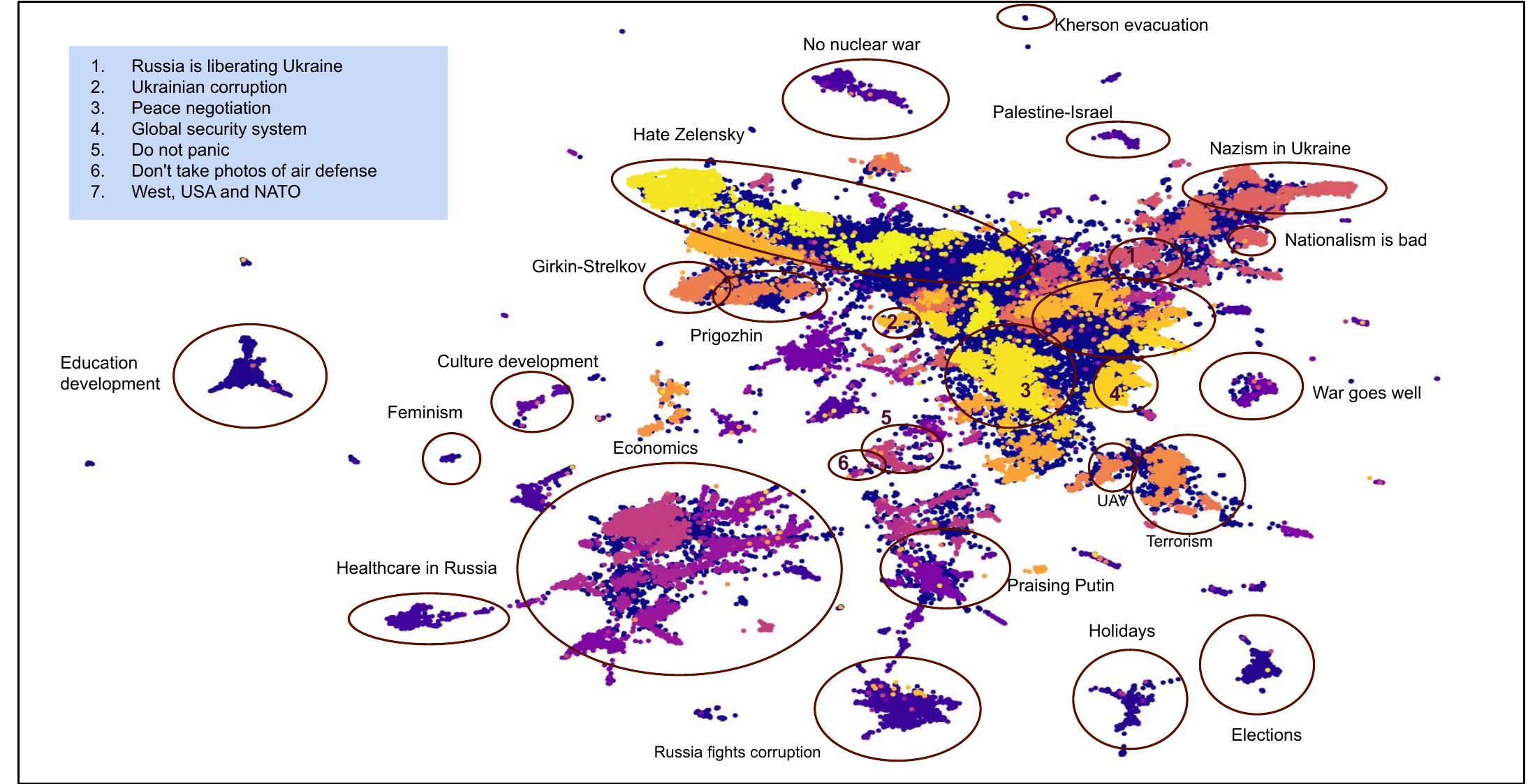}
    \caption{\textbf{Cluster map for the \pams.} Cluster map is generated using UMAP. Different colors represent different clusters. Some clusters and groups of clusters are annotated in order to illustrate main topics and narratives. The central area consists of multiple clusters, mostly about the Russian-Ukrainian war, these small clusters are denoted with numbers.}
    \label{fig:clusters}
\end{figure*}

\begin{figure*}[h!]
     \centering
     \begin{subfigure}[b]{0.65\textwidth}
         \centering
         \includegraphics[width=\textwidth]{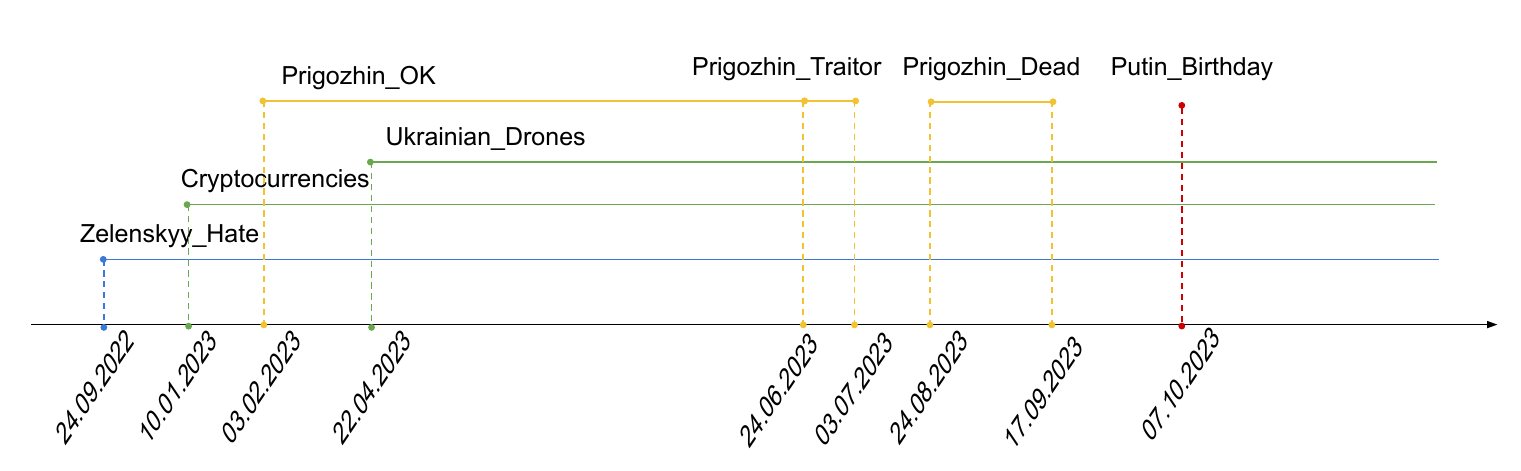}
         \label{fig:timeline}
     \end{subfigure}
     \hfill
     \begin{subfigure}[b]{0.3\textwidth}
         \centering
         \includegraphics[width=0.8\textwidth]{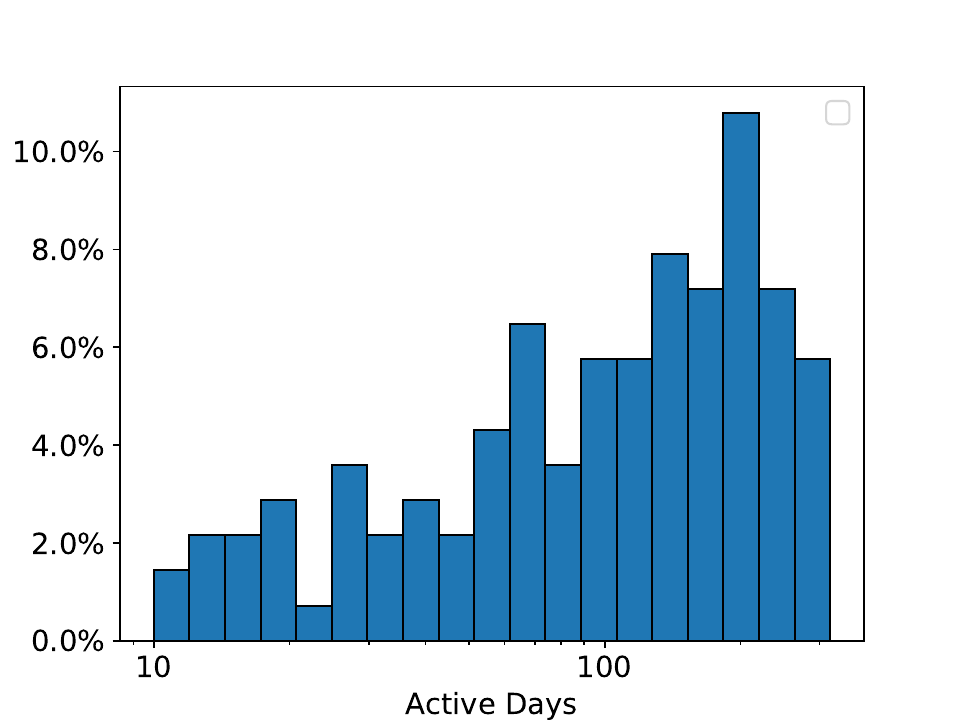}
     \end{subfigure}
     \hfill
    \caption{\textbf{Topics temporality}. \textit{Left: Timeline of Selected Topics} Note that the Prigozhin rebellion~\cite{wagner} started on the \nth{23} of June, but the corresponding messages appeared only the next day, on the 24th.\\ \textit{Right: Topic Longevity} - Total activity time for different topics in days. Almost half of the topics are ephemeral (their lifespan is less than 100 days).}
    \label{fig:timeline_hist}
\end{figure*}

\parabf{Topics} We now study the corpus of \pam texts from historical and real-time datasets to determine what narratives they spread. 
In order to identify \pas in the historical dataset, we use the same augmentation procedure as used for the real-time data (Section \ref{sec:data-label}). We obtain $\sim$60K unique messages.

We use a semi-automated approach to cluster topics. We apply DBSCAN to cluster SBERT~\cite{reimers-2019-sentence-bert} embeddings. We use the version of SBERT pre-trained on Russian language datasets\cite{ru_sbert}. Then, we augmented the resulting ~180 clusters manually: we searched for certain words, like "corruption" or "Zelensky", manually checked the texts, and assigned them to the corresponding clusters. In the end, we assign topics for $\sim$80\% of the \pams. Figure \ref{fig:clusters} illustrates the structure of the topic clusters in the 2D plane. The largest fraction of the corpus is made up of messages dedicated to the War in Ukraine. The second largest group of clusters are topics related to the internal policies of the Russian Federation. The topics can be divided into four broad groups:

\parait{Generic Propaganda} Narratives affine to the Russian Government's official agenda. The most popular are messages criticizing V. Zelensky and the Ukrainian government. A significant number of messages cover Russian domestic issues such as corruption, public healthcare, wages, demography, etc. This category also contains small topics like vaping, feminism, and cryptocurrency.

\example{Just look at Zelensky's behavior in public. He looks back and forth, nose sniffs, and hands don't find their place, it's the behavior of a typical junkie.}\\
    
\parait{Predictable Events} Topics associated with a certain date, usually a national holiday or a government-organized event such as elections. These include congratulatory messages or messages to promote engagement in statewide activities.

\example{Happy New Year to all citizens of Russia! I wish not to give up in the new year, to continue your journey to your dreams, and to achieve it.}\\

\parait{Unpredictable Events} Reactions to recent relevant events, which cannot be predicted. The reaction usually appears one or two days after the event (as illustrated in Fig. \ref{fig:timeline_hist}). Examples of such events are: the Wagner Group rebellion, the Israeli–Palestinian war in October 2022, the Armenia–Azerbaijan war escalation, or minor Russian internal events like the Moscow naked party in December 2023. 

\example{All Wagner's activities are illegal - if anyone wants to join them now, they become a traitor to their homeland.}\\

\parabf{Emotional reactions} - This group of topics contains emotional reactions to different messages. For example, reactions to criminal or accident news with condolences, despair, or support messages; or expressions of agreement on pro-Russian statements by users.


\example{{\normalfont (in response to a message about a murder that happened somewhere)} I cannot read this news. I feel sick when I imagine this picture in real life.}
    

\parabf{Topic Temporality} Now that we have an understanding of the types of topics, we consider their temporality. We observe that topic composition is not fixed over time (see Figure \ref{fig:timeline_hist}). Around 40\% of the topics persist over the entire observation period, while ~20\% of topics, typically associated with events, are active for short periods of time, often less than one month. We illustrate these shifts for some selected topics in Figure \ref{fig:timeline_hist}.






\section{Propaganda Detection}
\label{sec:detect}
In previous sections, we demonstrated that Telegram channels contain a large number of propaganda accounts. In this section, we propose methods for detecting and preventing these propaganda activities.
\subsection{Human Propaganda Moderation}
\label{sec:human-moderators}
On Telegram, channel-level moderation can be performed by human moderators who detect and clean propaganda activities by banning \pas and deleting \pams. We identify deleted \pams by comparing the real-time and historical datasets. We use this observation to detect the presence of propaganda moderation in a channel and measure its effectiveness, as reported in Table \ref{tab:bot-mdata}. 
We measure the moderation effectiveness as the ratio of \pams deleted by moderators to the total number of labeled \pams in a channel. We see a large variance in moderation effectiveness, ranging from below 20\% (Rudoi) to more than 80\% of propaganda messages removed (RT, Nexta, and Shtefanov). Comparing these ratios with the total moderation rate, i.e., the total number of deleted messages divided by the total number of messages in a given channel, we see that in some channels, the moderation of \pams is much more aggressive than other messages. By contacting the moderation team of the Shtefanov channel (87\% of \pams deleted), we confirmed that their high success rate is due to their moderation policy's focus on the detection and deletion of \pams alongside a substantial human effort into checking every message.  

\begin{table}[h]
\centering
\caption{\textbf{Propaganda moderation effectiveness} measured as the ratio of deleted \pams to the total number of \pams in a channel. Higher percentages indicate more aggressive moderation. We also report the ratio of all deleted messages to the total number of messages in the channel.}
\begin{tabular}{l c c c} 

 Channel & Size & Propaganda  & Total \\
            & & Moderation & Moderation \\
 \hline \\
 RT  & 584K & 94.7\% & 9.3\%  \\
 Shtefanov & 281K & 87.5\%  & 15.2\% \\
 Nexta  & 824K & 84.1\% & 19.6\%  \\
 SamaraNews  & 5.0K & 64.1\% & 56.8\% \\
 KK  & 158K & 45.2\% & 13.3\% \\
 Readovka & 863K & 38.5\% & 16.5\% \\
 Topor  & 297K & 29.6\%  & 26.2\% \\
 Rudoi & 47.3K & 19.9\% & 7.9\%\\
\end{tabular}
\label{tab:bot-mdata}
\end{table}

We conclude that some Telegram channels have a strong interest in propaganda moderation, and they do so mostly using manual detection and deletion. Manual detection has several drawbacks. First, it requires a dedicated staff (either hired or volunteered). 
Second, moderators are not always online, and their reaction time is limited by their concentration and reading abilities.  
Third, a non-negligible portion of \pams (5-15\%) remain visible to the users and attract interactions from them. 
Finally, human moderators are exposed to propaganda content, which may result in psychological issues, similar to hate speech or violent content moderation~\cite{steiger2021psychological}.

\subsection{Automated Propaganda Detector}

Automating propaganda detection would remove the need for an online dedicated staff, as it provides a better detection rate more quickly and cheaply than human moderators. To remain more effective than human moderators over time, it should also be at least as robust to changes in propaganda behavior, e.g., associated with new topics.

In line with work on other platforms (e.g., X~\cite{mazza2019rtbust} and Reddit~\cite{redditTrolls}), we consider detection solutions in the form of a machine-learning classification model, which can be deployed by channel owners via the bot API on Telegram. 

To be deployable, the classification model must only use information available via the API, as that is the only information moderators have access to.
This renders useless approaches based on relations between different accounts (e.g., based on connections in the social graph graphs~\cite{graphbots1, graphbots2, graphbots3}) because account contacts or the list of channels a given account has joined are not available via the API. 
The data available for moderators consists of account information (First Name, Last Name, Username), message metadata (date and time, size), and message content (message text, trigger message text). Since the account information can be easily hidden or manipulated, we do not use it for detection purposes.



\newcommand{\shallow}{handcrafted features\xspace}
\newcommand{\Shallow}{Handcrafted features\xspace}
\newcommand{\Trigger}{Trigger embeddings\xspace}
\newcommand{\trigger}{trigger embeddings\xspace}
\newcommand{\reply}{propaganda embeddings\xspace}
\newcommand{\Reply}{Propaganda embeddings\xspace}
\newcommand{\TREnsemble}{Trigger-Propaganda emsemble\xspace}
\newcommand{\tREnsemble}{trigger-propaganda emsemble\xspace}
\newcommand{\TREmbeddings}{Trigger-Propaganda embeddings\xspace}
\newcommand{\tREmbeddings}{trigger-propaganda embeddings\xspace}

\vfill\null

\noindent We implement the following approaches: 

\parabf{\Shallow} We use handcrafted features computed on messages' metadata and content. Such features are widely used in the bot~\cite{ferrara2016rise,twitterbot2017}, trolls~\cite{redditTrolls}, and spam~\cite{yardi2010detecting, spamcampaign} detection literature. Concretely, we select the following textual features from~\cite{wang2015making}, which are often used in these fields: \textit{message length}, \textit{number of words}, \textit{number of URL links},
\textit{number of emojis},
\textit{number of exclamation marks},
\textit{number of question marks},
\textit{message time in seconds},
\textit{latency between message and reply in seconds}.
We use these features to train XGBoost~\cite{chen2016xgboost}, RandomForest~\cite{ho1995random}, LightGBM~\cite{zhang2019lightgbm}, Decision Tree, Logistic Regression, and DNN classifiers. We only report results for the XGBoost model, since it achieved the best performance on our tasks.

\parabf{\Reply}
The content of propaganda messages is different from user messages in terms of specific word frequency and style (see Sect. \ref{sec:activity-stats}).
Content-based detection in the literature is based on the use of n-grams\cite{pizarro2019using}, or textual embeddings~\cite{wei2019twitter, garcia2019empirical, kumar2021content}.
In this work, we use the latter since textual embeddings are an n-grams generalization. Concretely, we use the same pre-trained SBERT embeddings that we use in Sect.~\ref{sec:narr} for clustering as they show good performance on other Russian language NLP tasks. We try several classifiers, including XGBoost, trained on these embeddings, and report results on the best performing one: a simple 3-layer DNN.
 
\parabf{\Trigger}
Trigger messages are also different from user messages in terms of the specific word frequency (Sect. \ref{sec:activity-stats}). An additional advantage over \pams is that they cannot be manipulated by propaganda creators. We use the same embeddings and classifiers as for \pams.

\parabf{\TREnsemble}
We combine the results of the classifiers trained on \trigger and \reply in an ensemble. We take as output the rounded sum of the output of these two detectors.

\parabf{\TREmbeddings}
Using an ensemble does not capture any relationship between trigger messages and \pams. Yet, we know that these pairs often have different appearances than normal conversations(see Conversation \ref{conv:first_examples}). To capture this mismatch, we assume that the textual information is partially preserved in the embedding, and feed the same DNN-based classifier with the concatenation of embedding pairs of triggers and the corresponding propaganda replies. 

\begin{table*}[hbt!]
\caption{\textbf{Detection Performance.} For human moderators, we report the precision since we do not have false positive data for the accuracy estimation.}
\begin{center}
\begin{tabular}{l c c c}
 Method & Overall Accuracy & New Topics Accuracy & Validation \\
 \hline \\
 Human Moderators & 86.0\% & 81.2\% &- \\
 \Shallow & 83.8\% & 88.6\% & 80.6\% \\
 \Trigger & 79.0\% & 41.3\% & 54.0\% \\
 \Reply & 96.8\% & 89.4\% & 81.2\% \\
 \TREnsemble & 96.5\% & 77.5\% & 73.4\% \\
 \TREmbeddings & \textbf{97.4\%} &  \textbf{93.0\%} & \textbf{88.8\%} \\
\end{tabular}
\end{center}

\label{tab:final-table}
\end{table*}

\subsection{Evaluation}
We evaluate all approaches with respect to the requirements in Section \ref{sec:human-moderators}, and report the results in Table \ref{tab:final-table}.
We include the effectiveness of human moderators (see Table \ref{tab:bot-mdata}) for comparison. We report the average effectiveness over channels that have aggressive propaganda deletion policies, i.e., \emph{Nexta}, \emph{RT}, and \emph{Shtefanov}. Since we do not know the reasons for deletion by human moderators and thus we cannot evaluate false positives, we report precision.  

\subsubsection{Automated detection performance}
We first evaluate whether automated detection can obtain better performance than human moderators.


We evaluate the automated detection approaches by training the classifiers on messages collected between August 16 and September 18, 2023, and testing their performance on messages collected between September 18 and October 16. This separation splits the data to train and test evenly and mimics a realistic scenario in which moderators deploying the detector can only train on labeled data from the past. To ensure that the evaluation is fair in terms of accuracy, we create a balanced dataset in terms of \pams and users' messages.

The results of this evaluation (\nth{2} column in Table \ref{tab:final-table}) show that using \tREmbeddings, \tREnsemble, and \reply as input to the classifier outperforms human labelers. Among these, \tREmbeddings performs the best, closely followed by only using the content of \pams.  Notably, while the \tREnsemble and \tREmbeddings use the same input data, there is a large difference in terms of performance. We interpret that this is because explicitly capturing the relation between trigger messages and their replies is important for detection. Also, \tREnsemble does not improve over just using the \pams, indicating that the trigger carries little information for detection. 


\subsubsection{Performance on unseen topics}

Propaganda messages may refer to events or facts that are not present in the training period. In this section, we study how the different automated approaches perform in such circumstances. We observed five new topics on the test set:
\begin{enumerate}[noitemsep, nosep, wide]
    \item \textit{Road Development}: about the growth of the road network in Russia.
    \item \textit{Alcoholism}: about the decrease in alcohol consumption in Russia, thanks to the introduction of new laws.
    \item \textit{Putin Birthday}: happy birthday to Vladimir Putin (on October 7).
    \item \textit{Armenia-Azerbaijan}: about the Nagorno-Karabakh war and Armenia-Azerbaijan relations in general (from October 20 after the escalation that started on October 19).
    \item \textit{Palestine-Israel}: about the Israel-Hamas war (from October 9 after the events on October 7).
\end{enumerate}

The third column in Table \ref{tab:final-table} shows the average accuracy across these new topics. Using \tREmbeddings provides the best capability to adapt to new topics (93.0\%), followed by using \reply (89.4\%). Using the trigger messages yields very poor results, likely due to overfitting. Whether alone or in the ensemble, trigger messages reduce the performance of the detector considerably. This is because while trigger messages are generated by different users and can greatly vary in language or length, the \pams follow certain style patterns, and this style can be preserved in embedding. 

The performance of all automated approaches decreases on new topics, and so does the performance of human moderators (5\% decrease). We conjecture that human moderators also need to "learn" these new topics, to efficiently delete the \pams associated with them. 


\begin{table*}[h!]
\caption{\textbf{Worst topic accuracy.} Red numbers indicate that the topic is in the top 5 worst topics per detector. Bolding indicates the best detector for each topic. For example, the \nth{2} worst topic for the \tREmbeddings was Terrorism and the handcrafted features performed the best on that topic. For human moderators, we use  "-" when topics have less than 50 messages in their channels.}
\begin{center}
\begin{tabular}{l c c c c}
\label{tab:worst-topics}
 Topic  & Trigger-Propaganda emb. & Propaganda emb.  & \Shallow & Human Moderators \\
 \hline \\
 Roads Developing & {\color{red} 60.0\%} & {\color{red} 60.0\%} &  \textbf{74.0\%} & - \\
 Terrorism & {\color{red} 68.1\%} & {\color{red}66.7\%} & \textbf{73.9\%} & - \\
 Alcoholism & \textbf{\color{red}77.5\%} & {\color{red}64.1\%} & {\color{red}23.3\%} & -\\
 Holidays & \textbf{\color{red}77.7\%} & {\color{red}40.7\%} & 63.0\% & - \\
 Education Developing & {\color{red}79.7\%} & \textbf{80.5\%} & 63.2\% & {\color{red}72.5\%} \\
 Sadness Emotion & \textbf{89.9\%} & {\color{red}69.5\%} & {\color{red}1.2\%} & 81.0\% \\
 Cryptocurrencies & 91.1\% & \textbf{97.0\%} & {\color{red}11.8\%} & - \\
 Sad News Emotion & \textbf{97.2\%} & 81.7\% & {\color{red}23.4\%} & 82.4\%\\
 Despair Emotion & \textbf{96.5\%} & 91.2\% & {\color{red}45.6\%} & 82.7\% \\
 Ukrainian Refugees & \textbf{100.0\%} & \textbf{100.0\%} & 98.9\% & {\color{red}77.7\%} \\
 Palestine-Israel & \textbf{88.8\%} & 81.5\% & 81.9\% & {\color{red}77.7\%}  \\
 Russia Helps & \textbf{99.3\%} & 97.8\% & \textbf{98.5\%} & {\color{red}77.8\%} \\
 Culture Developing & 99.3\% & \textbf{99.7\%} & 75.3\% & {\color{red}79.4\%} \\
\end{tabular}
\end{center}
\end{table*}

\subsubsection{Error analysis}
We now study whether detection performance degradation is due to the appearance of new topics or if certain topics are inherently more difficult to classify than others. 

\begin{figure}
    \centering
    \includegraphics[width=1\linewidth]{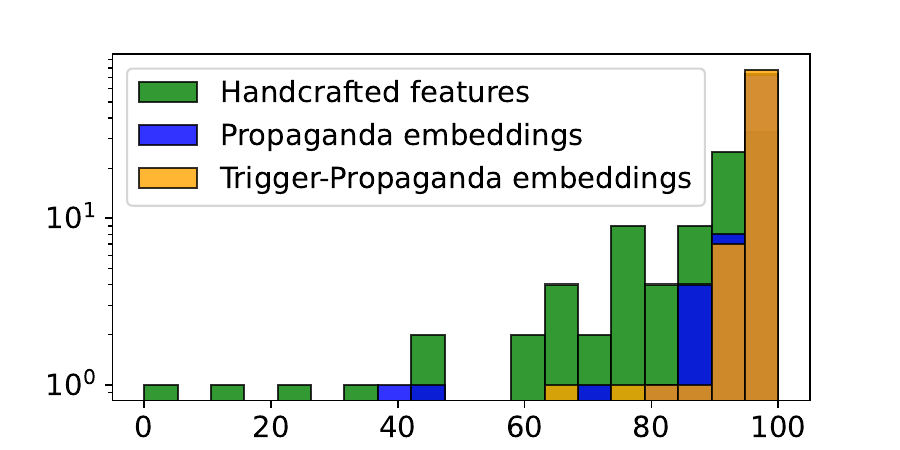}
    \caption{\textbf{Accuracy distribution for different topics.} \TREmbeddings demonstrate the most consistent performance across topics. \Shallow offer very bad performance on some topics.}
    \label{fig:score_vs_train}
\end{figure}

We plot the distribution of accuracy over topics of \shallow, \reply, and \tREmbeddings in Figure \ref{fig:score_vs_train}. Using \tREmbeddings demonstrates the most consistent results across topics. Using \shallow results in poor generalization, with some topics being particularly difficult to identify even if they exist in the training set.

We explore this issue in depth in Table \ref{tab:worst-topics}, where we show the performance of all approaches for the worst five topics for each detector. When the topics are present in moderated channels, we also report human moderators' performance. Using \shallow results in poor performance for emotional topics (e.g., sadness), cryptocurrency-related messages, and alcoholism. 
This is because message length is one of the main features used by the classifier based on \shallow, due to user messages being generally shorter than \pas. Thus, it performs poorly for topics where \pams are also short, e.g., cryptocurrency-related messages such as \textit{"I do not understand what crypto is."}.

We observe that the \textit{Holidays} topic is hard for all approaches. This is not surprising since holiday-related messages, such as \textit{``Happy New Year!'' } also appear in the user messages. Despite the difficulty of these topics, sometimes using \tREmbeddings can yield good results when the \pam is inconsistent with the trigger message (see  
Conversation \ref{conv:model_error_examples}). 


\begin{Conversation*}[h]
\centering
\begin{minipage}{0.4\textwidth}
\begin{chat}
\user{User1} One may ask, What does Putin’s birthday have to do with it?
\chatbot{"daniil" (ahanthuda)} Vladimir Vladimirovich, Happy Birthday! May everything go well for you!
\end{chat}
\end{minipage}
\hfill
\begin{minipage}{0.5\textwidth}
\begin{chat}
\user{User2} Happy 71st birthday to Putin!
He was born in Leningrad in 1952 on October 7, but despite his age, V. Putin is still as handsome as ever!
\chatbot{"Mark" (xiverelaroya)} Our leader is strong! I wish you a happy birthday, Vladimir Vladimirovich! 
\end{chat}
\end{minipage}
\vspace{0.2cm}
\caption{\textbf{Examples of errors for \reply detector (left) and for both \reply and \TREmbeddings detectors  (right)} Left: A \pa does not catch the irony and provides an unconnected reply. the Trigger-Reply system spotted the mismatch between the user message and reply, while the detection system using only the message information failed. 
Right: The conversation is completely normal, the reply matches the message, and even human labelers cannot detect a \pa based on this conversation.\\}
\label{conv:model_error_examples}
\end{Conversation*}

\begin{figure}[hbt!]
    \centering
    \includegraphics[width=1\linewidth]{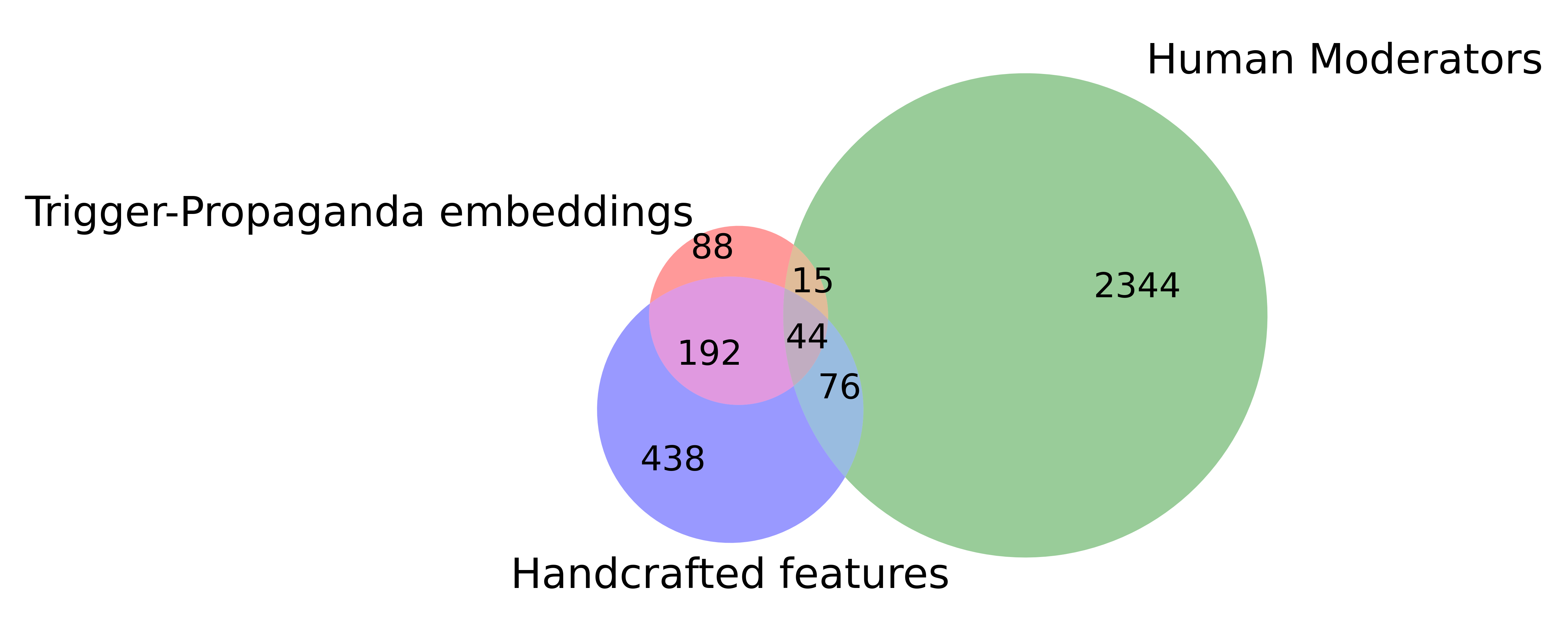}
    \caption{\textbf{False negatives in ML-based detectors and human moderators.} Most of the \tREmbeddings errors are also errors of \shallow, making it almost strictly superior.} 
    \label{fig:venn_diag}
\end{figure}

Last, we study whether human moderators make the same errors as automated detectors. We plot the intersection of human errors and \shallow and \TREmbeddings in Figure \ref{fig:venn_diag}. While most of the errors of the \TREmbeddings are shared with the \shallow, these errors are noticeably different from human moderation. In Table \ref{tab:worst-topics} we see that most topics that are problematic for humans are not hard to detect for ML-based detectors. In general, we could not find an explanation for why some topics are easier than others for humans. 
An interesting case is \textit{Putin\_Birthday} topic where human moderators have a precision of 94.2\% while the best ML-based detector (\tREmbeddings) only achieves a 91.5\% accuracy. 






\subsubsection{Validation on different propaganda accounts}
\label{sec:ukr-bots}

Although automated detection can generalize to unseen topics, it is still unclear if the detection methods can retain their performance if \pas change their behavior more radically. In this section, we evaluate the performance of the detectors on a second network that we discovered during our evaluation. 
\vfill\null
\parabf{Pro-Ukrainian propaganda network}
 When analyzing errors, we noticed that some of the false positives are messages with clear propaganda purposes but different content and account behavior. These messages contain pro-Ukrainian propaganda targeted at the Russian-speaking audience. The accounts writing them
repeat the messages from each other but never repeat the messages of the network we study in previous sections, which means that they form a different network. We call this network \textit{pro-Ukrainian}, as opposed to the \textit{pro-Russian} network studied in the previous sections. We missed these accounts during our initial labeling because they were not present in the channels we used for manual labeling ("Ru2ch", "Readovka"). 




The topics and style of the pro-Ukranian \pas are different from messages used by the pro-Russian network.
Pro-Ukranian accounts hide their usernames and their first and last names are notably different from the pro-Russian accounts: instead of common names, these accounts use completely fictional nicknames, e.g., "Atlanta" or "Az Air." The activity of these \pas is sporadic: they are active for short periods (1-2 days), disappear (20-30 days), and then reappear. Unlike the pro-Russian accounts, we also observe that they use reactions (e.g., likes) under each other's comments. 
Despite these differences, pro-Ukrainian accounts, as pro-Russian ones, often post replies unconnected with the trigger messages. We report more information about this network in Appendix \ref{app:ukr-network}.

\parabf{Evaluation} To evaluate automated detection on the pro-Ukrainian network, we repeat the labeling and augmentation from Sect. \ref{sec:data-label}  and obtain 2.7K \pams, which we balance with an equal number of user messages. 

We observe a performance degradation for all approaches (see Table~\ref{tab:final-table}, \nth{4} column). The embeddings-based methods show the most significant drop (11-15\%), while using \shallow results in a 6\% drop. This is because its main heuristic, the message length, remains useful for long pro-Ukrainian messages. Yet, using \tREmbeddings still provides the best accuracy (88.8\%), which we believe is due to its capacity to capture the relationship between triggers and replies.


\subsection{Deployment Considerations}
We now assess the financial and computational requirements associated with using \tREmbeddings, the best-performing detection approach. Since this detector includes two neural networks, it can be executed purely on CPU or with GPU acceleration. Renting a dedicated server with a CPU is much cheaper, which can be important for small channels that are not monetized.

We measure the average time for processing trigger-reply pairs one by one in the test set, which gives a worst-case timing estimation with respect to using badges. Using an NVIDIA RTX 3070 GPU, the average computation time is 0.015 $\pm$ 0.001 sec, while for an AMD Rysen 4700G CPU, the computation time increases to 0.25 $\pm$ 0.01 sec.
We do not have the technical means to measure the reaction time of human moderators (the Telegram API for deletion events is considered unreliable~\cite{telethon}). However, to our best knowledge, the visual reaction time for a human is more than 0.2 seconds ~\cite{amini2019evaluation}, without accounting for time to read, process the content, make a decision, and click all the buttons in the app (and the fact that humans cannot always be online). In fact, during manual labeling, human labelers could not label a message faster than in 1-3 seconds. We conclude that even using a slow CPU-based detector would result in a reaction time gain over human moderation.

Renting a GPU can even be profitable: a GPU node on Amazon AWS costs 0.21\$/h~\cite{amazon_gpu}, while the federal minimum wage in Russia is equal to 1.2\$/h, and the average salary is ~4\$/h~\cite{rus_salary}. In reality, a dedicated GPU for detection is unlikely to be fully loaded (due to the low frequency of incoming messages), and the GPU price can be further optimized using services like inference-on-demand~\cite{amazon_inference}. 

\section{Conclusion}
Telegram and other instant messengers are main sources of information in critical situations, in particular beyond the Western world (as in the case we study in this paper).
Our work evidences that due to its instant-messaging nature -- which is structurally different than the typically studied platforms in the literature such as X -- working on Telegram requires developing new collection and analysis methods.

Our Telegram-tailored collection and analysis allowed us to discover two large coordinated networks spreading propaganda and misinformation around the Russo-Ukrainian war and other politically-charged topics. 

The different nature of Telegram, where moderators are channel owners who have access to scarce information versus platform moderation, also forced us to design fully novel detection methods. We leveraged textual embeddings to capture the behavior of the \pas we found to obtain a quick and effective detector that improves over human moderators by a significant margin (11.6\%) and is robust to topic changes.

While future work should test to what extent our detection method generalizes to other propaganda campaigns on Telegram, this paper already shows that it is possible to help mitigate the threat of information-based attacks in instant messaging-based social networks. We hope that our results inspire the security community to broaden its attention beyond Western-centered social networks and build more tools to reduce information-based attacks worldwide. 



\bibliographystyle{IEEEtran}
\begin{small}
\bibliography{bib}

\begin{thebibliography}{10}
\providecommand{\url}[1]{#1}
\csname url@samestyle\endcsname
\providecommand{\newblock}{\relax}
\providecommand{\bibinfo}[2]{#2}
\providecommand{\BIBentrySTDinterwordspacing}{\spaceskip=0pt\relax}
\providecommand{\BIBentryALTinterwordstretchfactor}{4}
\providecommand{\BIBentryALTinterwordspacing}{\spaceskip=\fontdimen2\font plus
\BIBentryALTinterwordstretchfactor\fontdimen3\font minus
  \fontdimen4\font\relax}
\providecommand{\BIBforeignlanguage}[2]{{%
\expandafter\ifx\csname l@#1\endcsname\relax
\typeout{** WARNING: IEEEtran.bst: No hyphenation pattern has been}%
\typeout{** loaded for the language `#1'. Using the pattern for}%
\typeout{** the default language instead.}%
\else
\language=\csname l@#1\endcsname
\fi
#2}}
\providecommand{\BIBdecl}{\relax}
\BIBdecl

\bibitem{bradshaw2018challenging}
S.~Bradshaw and P.~N. Howard, ``Challenging truth and trust: A global inventory
  of organized social media manipulation,'' \emph{The computational propaganda
  project}, vol.~1, pp. 1--26, 2018.

\bibitem{times_hamas}
J.~Donovan, ``Misinformation is warfare,'' \emph{Time Magazine}, 2023,
  available at:
  \url{https://time.com/6323387/misinformation-israel-hamas-war-essay/}
  (Accessed: {May 30th, 2024}).

\bibitem{times_telegram}
V.~Bergenguen, ``How telegram became the digital battlefield in the
  russia-ukraine war,'' \emph{Time Magazine}, 2022, available at:
  \url{https://time.com/6158437/telegram-russia-ukraine-information-war/}
  (Accessed: {May 29th, 2024}).

\bibitem{ukraine_telegram}
\BIBentryALTinterwordspacing
USAID. (2023) Ukrainians increasingly rely on telegram channels for news and
  information during wartime. [Online]. Available:
  \url{https://internews.org/ukrainians-increasingly-rely-on-telegram-channels-for-news-and-information-during-wartime/}
\BIBentrySTDinterwordspacing

\bibitem{russia_telegram}
\BIBentryALTinterwordspacing
RE:RUSSIA. (2023) In russia, telegram has become the primary internet platform
  for young people, surpassing youtube in reach and whatsapp in average daily
  user time. [Online]. Available: \url{https://re-russia.net/en/review/263/}
\BIBentrySTDinterwordspacing

\bibitem{wijermars2022telegram}
M.~Wijermars and T.~Lokot, ``Is telegram a “harbinger of freedom”? the
  performance, practices, and perception of platforms as political actors in
  authoritarian states,'' \emph{Post-Soviet Affairs}, vol.~38, no. 1-2, pp.
  125--145, 2022.

\bibitem{twitterbot2017}
Z.~Gilani, E.~Kochmar, and J.~Crowcroft, ``Classification of twitter accounts
  into automated agents and human users,'' in \emph{2017 IEEE/ACM International
  Conference on Advances in Social Networks Analysis and Mining (ASONAM)},
  2017, pp. 489--496.

\bibitem{redditTrolls}
M.~H. Saeed, S.~Ali, J.~Blackburn, E.~De~Cristofaro, S.~Zannettou, and
  G.~Stringhini, ``Trollmagnifier: Detecting state-sponsored troll accounts on
  reddit,'' in \emph{2022 IEEE Symposium on Security and Privacy (SP)}.\hskip
  1em plus 0.5em minus 0.4em\relax IEEE, 2022.

\bibitem{zannettou2019disinformation}
S.~Zannettou, T.~Caulfield, E.~De~Cristofaro, M.~Sirivianos, G.~Stringhini, and
  J.~Blackburn, ``Disinformation warfare: Understanding state-sponsored trolls
  on twitter and their influence on the web,'' in \emph{Companion proceedings
  of the 2019 world wide web conference}, 2019.

\bibitem{hanley2023specious}
H.~W. Hanley, D.~Kumar, and Z.~Durumeric, ``Specious sites: Tracking the spread
  and sway of spurious news stories at scale,'' \emph{45th IEEE Symposium on
  Security and Privacy}, 2024.

\bibitem{hurtado2019bot}
S.~Hurtado, P.~Ray, and R.~Marculescu, ``Bot detection in reddit political
  discussion,'' in \emph{Proceedings of the fourth international workshop on
  social sensing}, 2019.

\bibitem{kumar2021content}
S.~Kumar, S.~Garg, Y.~Vats, and A.~S. Parihar, ``Content based bot detection
  using bot language model and bert embeddings,'' in \emph{2021 5th
  International Conference on Computer, Communication and Signal Processing
  (ICCCSP)}.\hskip 1em plus 0.5em minus 0.4em\relax IEEE, 2021, pp. 285--289.

\bibitem{rogers_telegram}
\BIBentryALTinterwordspacing
R.~Rogers, ``Deplatforming: Following extreme internet celebrities to telegram
  and alternative social media,'' \emph{European Journal of Communication},
  vol.~35, no.~3, pp. 213--229, 2020. [Online]. Available:
  \url{https://doi.org/10.1177/0267323120922066}
\BIBentrySTDinterwordspacing

\bibitem{rosebot}
\BIBentryALTinterwordspacing
P.~Larsen. (2023) Rosebot. [Online]. Available: \url{https://missrose.org}
\BIBentrySTDinterwordspacing

\bibitem{yayla2017telegram}
A.~S. Yayla and A.~Speckhard, ``Telegram: The mighty application that isis
  loves,'' \emph{International Center for the Study of Violent Extremism},
  vol.~9, 2017.

\bibitem{solopova2023automated}
V.~Solopova, O.-I. Popescu, C.~Benzm{\"u}ller, and T.~Landgraf, ``Automated
  multilingual detection of pro-kremlin propaganda in newspapers and telegram
  posts,'' \emph{Datenbank-Spektrum}, vol.~23, no.~1, pp. 5--14, 2023.

\bibitem{bovet2019influence}
A.~Bovet and H.~A. Makse, ``Influence of fake news in twitter during the 2016
  us presidential election,'' \emph{Nature communications}, vol.~10, no.~1,
  p.~7, 2019.

\bibitem{robles2022negativity}
J.-M. Robles, J.-A. Guevara, B.~Casas-Mas, and D.~G{\'o}mez, ``When negativity
  is the fuel. bots and political polarization in the covid-19 debate,''
  \emph{Comunicar}, vol.~30, no.~71, pp. 63--75, 2022.

\bibitem{baumgartner2020pushshift}
J.~Baumgartner, S.~Zannettou, M.~Squire, and J.~Blackburn, ``The pushshift
  telegram dataset,'' in \emph{Proceedings of the international AAAI conference
  on web and social media}, vol.~14, 2020, pp. 840--847.

\bibitem{Louvain}
V.~D. Blondel, J.-L. Guillaume, R.~Lambiotte, and E.~Lefebvre, ``Fast unfolding
  of communities in large networks,'' \emph{Journal of Statistical Mechanics:
  Theory and Experiment}, vol. 2008, p. P1000, 2008.

\bibitem{foreign_agent}
\BIBentryALTinterwordspacing
Wikipedia. (2024) Russian foreign agent law. [Online]. Available:
  \url{https://en.wikipedia.org/wiki/Russian_foreign_agent_law}
\BIBentrySTDinterwordspacing

\bibitem{luceri2019red}
L.~Luceri, A.~Deb, A.~Badawy, and E.~Ferrara, ``Red bots do it better:
  Comparative analysis of social bot partisan behavior,'' in \emph{Companion
  proceedings of the 2019 world wide web conference}, 2019, pp. 1007--1012.

\bibitem{varol2017online}
O.~Varol, E.~Ferrara, C.~Davis, F.~Menczer, and A.~Flammini, ``Online human-bot
  interactions: Detection, estimation, and characterization,'' in
  \emph{Proceedings of the international AAAI conference on web and social
  media}, 2017.

\bibitem{yang2013empirical}
C.~Yang, R.~Harkreader, and G.~Gu, ``Empirical evaluation and new design for
  fighting evolving twitter spammers,'' \emph{IEEE Transactions on Information
  Forensics and Security}, 2013.

\bibitem{beskow2018bot}
D.~M. Beskow and K.~M. Carley, ``Bot-hunter: a tiered approach to detecting \&
  characterizing automated activity on twitter,'' in \emph{Conference paper.
  SBP-BRiMS: International Conference on Social Computing, Behavioral-Cultural
  Modeling and Prediction and Behavior Representation in Modeling and
  Simulation}, 2018.

\bibitem{yang2020scalable}
K.-C. Yang, O.~Varol, P.-M. Hui, and F.~Menczer, ``Scalable and generalizable
  social bot detection through data selection,'' in \emph{Proceedings of the
  AAAI Conference on Artificial Intelligence}, 2020.

\bibitem{howard}
P.~N. Howard and B.~Kollanyi, ``Bots,\# strongerin, and\# brexit: Computational
  propaganda during the uk-eu referendum,'' \emph{Available at SSRN 2798311},
  2016.

\bibitem{dutta2020hawkeseye}
H.~S. Dutta, V.~R. Dutta, A.~Adhikary, and T.~Chakraborty, ``Hawkeseye:
  Detecting fake retweeters using hawkes process and topic modeling,''
  \emph{IEEE Transactions on Information Forensics and Security}, 2020.

\bibitem{retweets}
T.~Elmas, R.~Overdorf, and K.~Aberer, ``Characterizing retweet bots: The case
  of black market accounts,'' in \emph{Proceedings of the International AAAI
  Conference on Web and Social Media}, vol.~16, 2022, pp. 171--182.

\bibitem{gallwitz}
F.~Gallwitz and M.~Kreil, ``The rise and fall of `social bot' research,''
  \emph{SSRN: https://ssrn.com/abstract=3814191}, 2021.

\bibitem{wagner}
\BIBentryALTinterwordspacing
Wikipedia. (2024) Wagner group rebellion. [Online]. Available:
  \url{https://en.wikipedia.org/wiki/Wagner_Group_rebellion}
\BIBentrySTDinterwordspacing

\bibitem{reimers-2019-sentence-bert}
\BIBentryALTinterwordspacing
N.~Reimers and I.~Gurevych, ``Sentence-bert: Sentence embeddings using siamese
  bert-networks,'' in \emph{Proceedings of the 2019 Conference on Empirical
  Methods in Natural Language Processing}.\hskip 1em plus 0.5em minus
  0.4em\relax Association for Computational Linguistics, 11 2019. [Online].
  Available: \url{http://arxiv.org/abs/1908.10084}
\BIBentrySTDinterwordspacing

\bibitem{ru_sbert}
\BIBentryALTinterwordspacing
SberDevices. (2022) Bert large model (uncased) for sentence embeddings in
  russian language. [Online]. Available:
  \url{https://huggingface.co/ai-forever/sbert_large_nlu_ru}
\BIBentrySTDinterwordspacing

\bibitem{steiger2021psychological}
M.~Steiger, T.~J. Bharucha, S.~Venkatagiri, M.~J. Riedl, and M.~Lease, ``The
  psychological well-being of content moderators: the emotional labor of
  commercial moderation and avenues for improving support,'' in
  \emph{Proceedings of the 2021 CHI conference on human factors in computing
  systems}, 2021, pp. 1--14.

\bibitem{mazza2019rtbust}
M.~Mazza, S.~Cresci, M.~Avvenuti, W.~Quattrociocchi, and M.~Tesconi, ``Rtbust:
  Exploiting temporal patterns for botnet detection on twitter,'' in
  \emph{Proceedings of the 10th ACM conference on web science}, 2019, pp.
  183--192.

\bibitem{graphbots1}
S.~Feng, Z.~Tan, H.~Wan, N.~Wang, Z.~Chen, B.~Zhang, Q.~Zheng, W.~Zhang,
  Z.~Lei, S.~Yang \emph{et~al.}, ``Twibot-22: Towards graph-based twitter bot
  detection,'' \emph{Advances in Neural Information Processing Systems},
  vol.~35, pp. 35\,254--35\,269, 2022.

\bibitem{graphbots2}
S.~Ali~Alhosseini, R.~Bin~Tareaf, P.~Najafi, and C.~Meinel, ``Detect me if you
  can: Spam bot detection using inductive representation learning,'' in
  \emph{Companion proceedings of the 2019 world wide web conference}, 2019, pp.
  148--153.

\bibitem{graphbots3}
S.~H. Moghaddam and M.~Abbaspour, ``Friendship preference: Scalable and robust
  category of features for social bot detection,'' \emph{IEEE Transactions on
  Dependable and Secure Computing}, vol.~20, no.~2, pp. 1516--1528, 2022.

\bibitem{ferrara2016rise}
E.~Ferrara, O.~Varol, C.~Davis, F.~Menczer, and A.~Flammini, ``The rise of
  social bots,'' \emph{Communications of the ACM}, 2016.

\bibitem{yardi2010detecting}
S.~Yardi, D.~Romero, G.~Schoenebeck \emph{et~al.}, ``Detecting spam in a
  twitter network,'' \emph{First Monday}, 2010.

\bibitem{spamcampaign}
Z.~Chu, I.~Widjaja, and H.~Wang, ``Detecting social spam campaigns on
  twitter,'' in \emph{International Conference on Applied Cryptography and
  Network Security}, 2012.

\bibitem{wang2015making}
B.~Wang, A.~Zubiaga, M.~Liakata, and R.~Procter, ``Making the most of
  tweet-inherent features for social spam detection on twitter,'' in
  \emph{Proceedings of the the 5th Workshop on Making Sense of Microposts
  co-located with the 24th International World Wide Web Conference {(WWW}
  2015), Florence, Italy, May 18th, 2015}, ser. {CEUR} Workshop Proceedings,
  2015.

\bibitem{chen2016xgboost}
T.~Chen and C.~Guestrin, ``Xgboost: A scalable tree boosting system,'' in
  \emph{Proceedings of the 22nd acm sigkdd international conference on
  knowledge discovery and data mining}, 2016, pp. 785--794.

\bibitem{ho1995random}
T.~K. Ho, ``Random decision forests,'' in \emph{Proceedings of 3rd
  international conference on document analysis and recognition}, vol.~1.\hskip
  1em plus 0.5em minus 0.4em\relax IEEE, 1995, pp. 278--282.

\bibitem{zhang2019lightgbm}
J.~Zhang, D.~Mucs, U.~Norinder, and F.~Svensson, ``Lightgbm: An effective and
  scalable algorithm for prediction of chemical toxicity--application to the
  tox21 and mutagenicity data sets,'' \emph{Journal of chemical information and
  modeling}, vol.~59, no.~10, pp. 4150--4158, 2019.

\bibitem{pizarro2019using}
J.~Pizarro, ``Using n-grams to detect bots on twitter.'' in \emph{CLEF (Working
  Notes)}, 2019.

\bibitem{wei2019twitter}
F.~Wei and U.~T. Nguyen, ``Twitter bot detection using bidirectional long
  short-term memory neural networks and word embeddings,'' in \emph{2019 First
  IEEE International conference on trust, privacy and security in intelligent
  systems and applications (TPS-ISA)}.\hskip 1em plus 0.5em minus 0.4em\relax
  IEEE, 2019, pp. 101--109.

\bibitem{garcia2019empirical}
\BIBentryALTinterwordspacing
A.~Garcia-Silva, C.~Berrio, and J.~M. G{\'o}mez-P{\'e}rez, ``An empirical study
  on pre-trained embeddings and language models for bot detection,'' in
  \emph{Proceedings of the 4th Workshop on Representation Learning for NLP
  (RepL4NLP-2019)}, I.~Augenstein, S.~Gella, S.~Ruder, K.~Kann, B.~Can,
  J.~Welbl, A.~Conneau, X.~Ren, and M.~Rei, Eds.\hskip 1em plus 0.5em minus
  0.4em\relax Florence, Italy: Association for Computational Linguistics, Aug.
  2019, pp. 148--155. [Online]. Available:
  \url{https://aclanthology.org/W19-4317}
\BIBentrySTDinterwordspacing

\bibitem{telethon}
\BIBentryALTinterwordspacing
Telethon. (2024) Telethon documentation. [Online]. Available:
  \url{https://docs.telethon.dev/en/latest/modules/events.html}
\BIBentrySTDinterwordspacing

\bibitem{amini2019evaluation}
R.~Amini~Vishteh, A.~Mirzajani, E.~Jafarzadehpour, and S.~Darvishpour,
  ``Evaluation of simple visual reaction time of different colored light
  stimuli in visually normal students,'' \emph{Clinical Optometry}, pp.
  167--171, 2019.

\bibitem{amazon_gpu}
\BIBentryALTinterwordspacing
Amazon. (2024) Amazon ec2 g4 instances. [Online]. Available:
  \url{https://aws.amazon.com/ec2/instance-types/g4/}
\BIBentrySTDinterwordspacing

\bibitem{rus_salary}
\BIBentryALTinterwordspacing
F.~S.~S. Service. (2023) Accrued average nominal wage and median wage grew by
  14.1
  \url{https://eng.rosstat.gov.ru/folder/13901/document/215855}
\BIBentrySTDinterwordspacing

\bibitem{amazon_inference}
\BIBentryALTinterwordspacing
Amazon. (2024) Amazon inference on demand. [Online]. Available:
  \url{https://aws.amazon.com/machine-learning/elastic-inference/}
\BIBentrySTDinterwordspacing

\end{thebibliography}
\end{small}

\newpage
\appendices
\appendices
  
\section{Real-Time Collection mechanism}
\label{app:real-client}
In order to collect data in real-time mode, we developed a simple custom Telegram client application using the telethon library~\cite{telethon}, which we officially registered as such on the Telegram website and received the required API key. The client application is deployed on the full-time running and protected server and operates in the following way: after an account logs in to the client, for all groups that this account has joined, the client receives events from the Telegram server when a new message appears in a group. The client checks if the source of the message reported in the event is in our list of observed groups, and if this is the case, the application saves the messages as a JSON string, which we later refine in order to maintain the same format as we have in the historical data. 
To log into this client, we use a Telegram account belonging to one of the authors. 
This method allows us to download group messages, channel posts, and comments, since every channel has the attached "Discussion" group, where all comments and posts are shown as messages.

\section{Conversation samples sources in Russian and Ukrainian}
\label{app:conv-samples}


\noindent Original version of the Conversation~\ref{conv:example_first}:

\begin{chat}
\user{User} \foreignlanguage{russian}{"Україна була і буде завжди свободна і незалежна ні від кого тільки рашисти уроди хотіли захватити за 3 дні а получили хуй в сраку ,вам пизда уже убивці Україна Переможе ! Слава Нашим Бійцям ! Слава Україні!"}
\chatbot{"Michelle Ortega" (venonisa)} \foreignlanguage{russian}{"Не один освобожденный житель украинских городов, где ведутся боевые действия, уже убедился в том, что Россия не пытается захватить Украину, а лишь освобождает её от нацистского давления, оказывающего невероятную опасность для людей России и Украины."}
\end{chat}


\noindent Original version of the Conversation~\ref{conv:first_examples} (Left):

\begin{chat}
\user{User1} \foreignlanguage{russian}{"Отдельный угар в том, что вербуют в зону сво прямо на некоторых военных предприятиях. То есть они готовы бросить в пламя войны даже самых необходимых в моменте специалистов"}
\chatbot{\foreignlanguage{russian}{"Лира Капустина"} (unknown username)}%
\foreignlanguage{russian}{"Нацбаты и ЧВК не имеют отношения к официальной армии. Их мало интересуют приказы официальных властей. В прямом смысле слова неуправляемые берсерки, вооруженные до зубов. Само по себе существование вот таких вот боевых отрядов в Украине одна из причин денацификации."}
\end{chat}

\noindent Original version of the Conversation~\ref{conv:first_examples} (Right):

\begin{chat}
\user{User2} \foreignlanguage{russian}{"Бля а мужчины выскочили как самые главные знатоки феменизма
Оставьте феменизм для женщин суки"}
\chatbot{"Gesha" (ronashisi)} \foreignlanguage{russian}{"Радфем - серьезная болезнь. Мое мнение не изменить."}
\end{chat}

\noindent Original version of the Conversation~\ref{conv:model_error_examples} (Left):
\begin{chat}
\user{User1} \foreignlanguage{russian}{"Казалось бы, при чем тут день рождения Путина "}
\chatbot{"daniil" (ahanthuda)} \foreignlanguage{russian}{"Владимир Владимирович, с Днем Рождения, пусть у Вас всё будет просто хорошо!"}
\end{chat}

\noindent Original version of the Conversation~\ref{conv:model_error_examples} (Right):

\begin{chat}
\user{User2} \foreignlanguage{russian}{"Поздравляем Путина с 71-м Днём рождения. Родился он в Ленинграде в 1952 году 7 октября, не смотря свой возраст В.Путин всё также красив как и всегда."}
\chatbot{"Mark" (xiverelaroya)} \foreignlanguage{russian}{"Насколько сильный у нас лидер! От всей души поздравляю вас, Владимир Владимирович!"}
\end{chat}

\section{GPT-4 prompts used in the paper}
\label{app:gpt}

\noindent Prompt used for the random-username experiment:

\vspace{1mm}

\texttt{After a string @@@, I will give you a username. Tell me please if this username contains a clear reference to something in Russian or English language. The reference can be to first or last names, events, movies, literature, history, nature, pop-culture, etc. If there is no reference, just answer one word 'No', otherwise say 'Yes' and explain the reference.
Note that users can replace some letters in usernames by digits, e.g. 'i' can be replaced with '1' or 'o' can be replaced with '0'.
@@@}

\vspace{2mm}

\noindent Prompt used for the western-username experiment:

\vspace{1mm}

\texttt{After a string @@@, I will give you a username. Tell me please if this username is a combination of the first name and the last name common for the United States or United Kingdom. Note, that the last name can have some additional numbers or characters at the end like in "Smith5" or "Smithk". If it is explain why, if it is not just output one word "No.". 
@@@}

\section{Selected topics}

In this Appendix, we list all the topics mentioned in the Section \ref{sec:detect} with brief descriptions and examples.

\parabf{Roads Developing} This topic contains messages explaining that the road system in Russia is constantly improved by the government. 

\example{Now it's very easy to solve the problem of dangerous sections of roads, pits and holes -- you just have to go through the State Service App (Gosuslugi) -- it's gonna be quick!}

\parabf{Terrorism} Messages explaining that the Russian government fights terrorism.

\example{It's great that in Russia, day and night, the government fights terrorists and other threats, providing security for the citizens of the country!}

\parabf{Alcoholism} Messages in this topic explain that the situation with alcohol consumption is improving in Russia, and the government has introduced working policies.

\example{Yeah, there are a lot of rehabilitation centres now, so there are fewer drunkies since they're going straight to treatment.}

\parabf{Holidays} Messages tied to certain national Holidays, such as New Year, Constitution Day, Mother's Day, etc. Example:

\example{I want to wish all of you a new year of fulfilling all your wishes, and all your dreams come true!}

\parabf{Education Development} This topic contains texts explaining that the Russian educational system is good and is improving every year. Example:

\example{I'm so happy that Russian education is now developing very dynamically. My sister is studying at Moscow State University school - she really likes it}

\parabf{Culture Developing} Similar to the Education Development, but about culture. 

\example{No one in Russia would neglect cultural development! We have so many talented people who fantasize about amazing ideas, and the state is helping to make this happen!}

\parabf{Sad News Emotion} These messages contain emotional responses to user messages containing information about crimes, disasters, etc.

\example{I wish there were less news like this
}

\example{I'm shocked by this kind of news
}

\parabf{Sadness Emotion} Similar to the previous topic, but not tied to the news, just expressing sadness.

\example{That's so fucking gross.}

\example{Fuck. Is it possible not to see something like that again?}

\parabf{Despair Emotion} Another emotional topic, more about fear. 

\example{Fuck, that's awful.}

\example{It's scary, so scary.}

\parabf{Putin Birthday} Messages wishing Putin a happy birthday. 

\example{Vladimir Vladimirovich is really working hard for Russia, he's doing a lot for us. Happy birthday, our president!}

\parabf{Cryptocurrencies} Messages expressing doubt about cryptocurrencies. 

\example{I don't think the crypto is gonna be anything serious, it's just a toy.}

\example{Crypto here, crypto there, and it is a fucking soap bubble which is hyped all over the place.}

\parabf{Income} Messages convincing people that the average income is not getting worse or that the government controls the process. Both personal examples and general statements. 

\example{Well, don't make it up, even if we've got a little lower income, the authorities are already keeping that matter under control.}

\example{I don't know who's earning less now. Personally, I'm fine.}

\parabf{Ukrainian Refugees} Messages explaining that Ukraine must stop the war if they want refugees to go back home. 

\example{In general, I understand that the refugee situation could have been avoided easily. Zelenskyy, if he were worried about the people, would have made a truce with Russia at the beginning. Now he has to do the same thing right now, so that more people don't run away to other countries.}

\parabf{Palestine-Israel} Messages regarding Israeli–Palestinian conflict. Interestingly, most of the messages were pushing towards immediate peace; also there are messages putting the blame for this conflict on the US. 

\example{It seems to me that the only way to resolve this whole situation between Palestine and Israel is through peace talks, other methods are not working.}

\example{The US can help in a peaceful solution, but they always pick up a scenario that only triggers a war: it was in Ukraine, now we're seeing it in Israel!}

\parabf{Russia Helps} Messages explaining that Russia helps common Ukrainians.  

\example{We are not going to leave people in the liberated towns and settlements; we are willing to continue to support them until the situation improves, and there are a lot of videos on the internet directly from those delivering humanitarian aid.}

\section{Pro-Ukrainian \pa network}
\label{app:ukr-network}

\begin{figure}[h!]
    \centering
    \includegraphics[width=\linewidth]{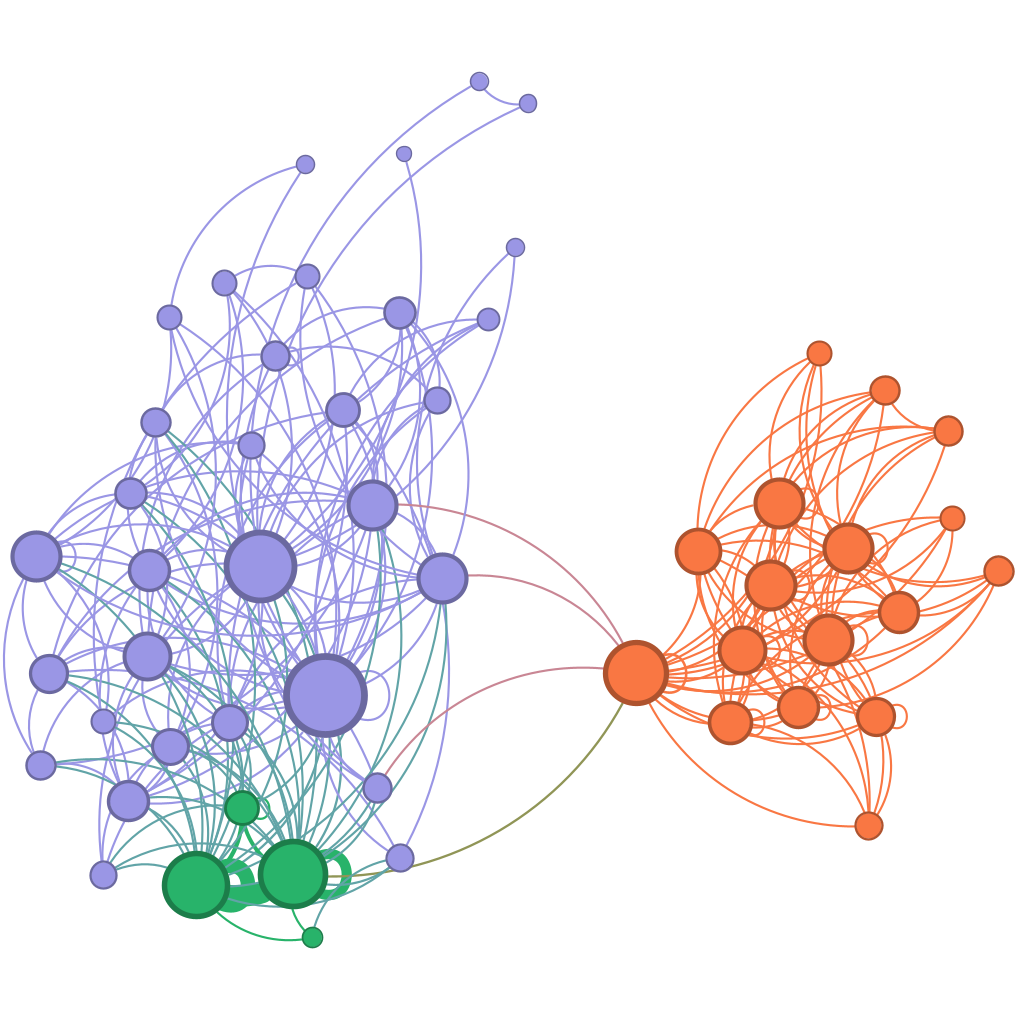}
    \caption{\textbf{Community structures for the pro-Ukrainian \pa network} Unlike pro-Russian network in Figure \ref{fig:coordination}, accounts form 3 distinctive communities, formed during sporadic activity periods. The green cluster appeared first on May 25-27th, 2023, followed by the purple one on July 1-7th, and the orange during short periods in August, September and October.}
    \label{fig:pro-ukr-coord}
\end{figure}

In this appendix, we give a brief analysis of our data regarding the pro-Ukrainian network introduced in Section \ref{sec:ukr-bots}. We have labeled 2.7K messages from 53 different accounts, operating from May 25 to October 5, 2023.

The examples of messages used by the pro-Ukrainian network include:

\example{"Under Putin's leadership, Russia has witnessed systematic violations of human rights, restrictions on freedom of speech, and the suppression of opposition."}

\example{"Ukraine has all the signs of a sovereign state: its own constitution, economy, army, and it represents the interests of its people in the international arena."}

\example{"The concentration of power in just one man's hand can slow down the decision-making process and lead to an insufficient response to challenges and changes in society and the world."}

On Figure \ref{fig:pro-ukr-coord}, we build the community graph using the same text-repetition 
 method  as in Section \ref{sec:coord}. Overall, it supports the hypothesis that these networks have different origins and behavior. While the pro-Russian network does not demonstrate distinctive communities, the pro-Ukrainian ones form three distinctive clusters, associated with short periods of their sporadic activity.

\end{document}